\documentclass[amsmath,amssymb,superscriptaddress,showpacs,twocolumn]{revtex4}
\usepackage{graphicx}
\usepackage{dcolumn}
\usepackage{amsmath}
\usepackage[utf8]{inputenc}
\usepackage{amssymb}
\usepackage{natbib}
\usepackage{booktabs}
\usepackage{multirow}
\usepackage{tabularray}
\usepackage{gensymb}
\usepackage{lipsum}
\usepackage{blindtext}
\usepackage{mathtools}
\usepackage{tensor}
\usepackage[colorlinks=true, citecolor=blue, urlcolor = blue, linkcolor= red, bookmarks=true]{hyperref}
\usepackage{float}
\usepackage{amsfonts}
\usepackage{pgfplots}
\usepackage{mathrsfs}
\usepackage{afterpage}

\begin{document}
\title[]{Investigating Rotating Black Holes in Bumblebee Gravity: Insights from EHT Observations}
\author{Shafqat Ul Islam} \email{Shafphy@gmail.com}
\affiliation{Astrophysics and Cosmology Research Unit, 
	School of Mathematics, Statistics and Computer Science, 
	University of KwaZulu-Natal, Private Bag 54001, Durban 4000, South Africa}
\author{Sushant~G.~Ghosh }\email{sghosh2@jmi.ac.in}
\affiliation{Centre for Theoretical Physics, 
	Jamia Millia Islamia, New Delhi 110025, India}
\affiliation{Astrophysics and Cosmology Research Unit, 
	School of Mathematics, Statistics and Computer Science, 
	University of KwaZulu-Natal, Private Bag 54001, Durban 4000, South Africa}
\author{Sunil D. Maharaj} \email{ maharaj@ukzn.ac.za}
\affiliation{Astrophysics and Cosmology Research Unit, 
	School of Mathematics, Statistics and Computer Science, 
	University of KwaZulu-Natal, Private Bag 54001, Durban 4000, South Africa}

\begin{abstract}
The EHT observation revealed event horizon-scale images of the supermassive black holes Sgr A* and M87* and these results are consistent with the shadow of a Kerr black hole as predicted by general relativity. However, Kerr-like rotating black holes in modified gravity theories can not ruled out, as they provide a crucial testing ground for these theories through EHT observations. It motivates us to investigate the Bumblebee theory, a vector-tensor extension of the Einstein-Maxwell theory that permits spontaneous symmetry breaking, resulting in the field acquiring a vacuum expectation value and introducing Lorentz violation. We present rotating black holes within this bumblebee gravity model, which includes an additional parameter $\ell$ alongside the mass $M$ and spin parameter $a$ -  namely RBHBG. Unlike the Kerr black hole, an extremal RBHBG, for $\ell<0$, refers to a black hole with angular momentum $a>M$. We derive an analytical formula necessary for the shadow of our rotating black holes, then visualize them with varying parameters $a$ and $\ell$, and also estimate the black hole parameters using shadow observables viz. shadow radius $R_s$,  distortion $\delta_s$, shadow area $A$ and oblateness $D$ using two well-known techniques. We find that $\ell$ incrementally increases the shadow size and causes more significant deformation while decreasing the event horizon area. Remarkably, an increase in $\ell$ enlarges the shadow radius irrespective of spin or inclination angle $\theta_0$. 
\end{abstract}
%\pacs{04.50.Kd, 04.20.Jb, 04.40.Nr, 04.70.Bw}
\maketitle
\section{Introduction}\label{Sec-1}
The Standard Model (SM) of particle physics and General Relativity (GR) are two fundamental theories describing the natural world: SM addresses particles and quantum interactions. In contrast, GR describes classical gravitation \citep{Griffiths:2008zz}. Unifying these theories is crucial for comprehensively understanding nature, leading to various proposed quantum gravity (QG) theories \citep{Rovelli:2004tv}. Directly testing QG is challenging due to the required Planck scale energies (~$10^{19}$ GeV), but potential signals, such as Lorentz symmetry breaking, might be detectable at lower energy scales.

Lorentz invariance, a fundamental assumption of GR, is a key symmetry verified with great precision \citep{Schutz:1985jx}. However, the potential for its violation remains a topic of active debate \citep{Liberati:2013xla,Mattingly:2005re}. This paper explores the implications of Lorentz symmetry breaking (LSB) in gravitation, which can be studied through the Standard Model Extension \citep{Kostelecky:2003fs}, incorporating a gravitational sector and Lorentz-violating terms \citep{Bluhm:2007xzd}. The idea of LSB is intriguing as it arises in various theoretical frameworks, including string theory \citep{Kostelecky:1988zi,Kostelecky:1989jp}, noncommutative field theories \citep{Carroll:2001ws}, and loop quantum gravity \citep{Gambini:1998it}.

One such simple model is Bumblebee gravity, where the vacuum expectation value of a vector field spontaneously breaks Lorentz symmetry ~\citep{Kostelecky:1989jw}. Bumblebee gravity black hole solutions and Lorentz violation effects have been actively investigated in recent years \citep{Kostelecky:1988zi,Bluhm:2004ep,Bertolami:2005bh,Bailey:2006fd,Bluhm:2008yt,Seifert:2009gi,Maluf:2014dpa,Paramos:2014mda,Assuncao:2019azw,Escobar:2017fdi}.  Casana \textit{et al} initially established an exact solution for a static, uncharged black hole and examined several classic investigations \citep{Casana:2017jkc}.  The black hole spacetime in Bumblebee gravity has been studied for gravitational lensing ~\citep{Ovgun:2018ran}, quasinormal modes \citep{Oliveira:2021abg} and  Hawking radiation \citep{Kanzi:2019gtu}.  Additionally, spherically symmetric black hole solutions with global monopole \citep{Gullu:2020qzu}, cosmological constant  \citep{Maluf:2020kgf}, Einstein-Gauss-Bonnet term \citep{Ding:2021iwv} and traversable wormhole solution  \citep{Ovgun:2018xys} have been discovered in this spacetime.  The cosmic consequences of the bumblebee gravity model are further explored in \citep{Capelo:2015ipa}.

Astrophysical objects have non-vanishing spin angular momentum; hence, black hole observational tests usually require the solution of spinning black holes.  The spinning black hole solution was found in Bumblebee gravity \citep{Ding:2019mal}, examining the effect of LSB parameter $\ell$ on the black hole shadow shape. No analysis was conducted to constrain the LSB parameter $\ell$ from theoretical predictions and observations of the supermassive black hole in M87* ~\citep{EventHorizonTelescope:2019dse}. However, future observations of black hole shadows may measure the parameter's value.  In this gravity, the shadow \citep{Ding:2019mal,Wang:2021irh}, accretion disk \citep{Liu:2019mls}, superradiant instability \citep{Jiang:2021whw}, and particle motion surrounding the black hole  \citep{Li:2020dln} are already examined.  These experiments help test Bumblebee theory and identify Lorentz symmetry breakdown from the vector field.

Moreover, constraints on the LSB parameter introduced by the bumblebee field have been rigorously investigated using a range of astrophysical data \citep{Casana:2017jkc,Wang:2021gtd,Gu:2022grg,Wang:2021irh}. Quasi-periodic oscillation (QPO) frequencies in X-ray emissions from black hole accretion discs have been utilized to place bounds on these parameters \citep{Wang:2021gtd}. QPOs, which reflect the dynamics and structure of the accretion disc, can reveal deviations from Lorentz invariance due to their sensitivity to the spacetime geometry around black holes. Studies leveraging the spectral data from the 2019 NuSTAR observation of the Galactic black hole EXO 1846-031 have provided crucial insights. The detailed analysis of X-ray emissions from this black hole's accretion disc allows researchers to infer the effects of Lorentz-violating fields on the observed frequencies, thus constraining the bumblebee field parameters \citep{Gu:2022grg}. Furthermore, the angular diameter of the shadow of the supermassive black hole M87*, as captured by the Event Horizon Telescope (EHT), has been another critical observational tool \citep{Wang:2021irh}. 

This paper considers a rotating metric in Bumblebee gravity that is slightly different and more straightforward than the one previously obtained \citep{Ding:2019mal}. Our analysis aims to impose more stringent constraints on the LSB parameter by utilizing the EHT results of shadow observables for both Sgr A* and M87*. We provide precise bounds on how deviations from Lorentz invariance influence the observed shadow characteristics. The high-resolution data from EHT enables a detailed comparison between observed and theoretical shadow profiles, revealing subtle effects of Lorentz-violating fields.   This approach enhances our understanding of fundamental deviations in astrophysical environments and refines constraints on quantum gravity theories. It may be helpful to provide deeper insights into the nature of spacetime at the Planck scale and the universe's underlying structure.

This paper is organized as follows: In Sec. \ref{Sec-2}, we briefly present new rotating black hole solutions within Bumblebee gravity, including an overview of their spacetime structure and parameter space. Sec. \ref{Sec-3} focuses on studying black hole shadows, emphasizing photon orbits and utilizing null geodesics. Sec. \ref{Sec-4} is dedicated to analyzing shadow observables and parameter estimation, with a detailed discussion on the influence of the LSB parameter on these observables. In Sec. \ref{Sec-5}, we constrain the LSB parameters based on M87* and Sgr A* observations. Finally, we summarize our findings and discuss their implications in Sec. \ref{Sec-6}. 

Throughout the paper, we adopt geometric units where $G=c=1$ unless stated otherwise.

\section{Black Holes in Bumblebee Gravity}\label{Sec-2}
We examine the bumblebee model, which extends GR by introducing a vector field with a non-vanishing vacuum expectation value, causing spontaneous Lorentz symmetry breaking \citep{Kostelecky:1989jw,Bluhm:2004ep}. First, we review Bumblebee gravity and derive a simpler rotating black hole solution than the one in Ref.~\citep{Ding:2019mal}.

\begin{figure}[t]
\begin{centering}
\includegraphics[scale=0.7]{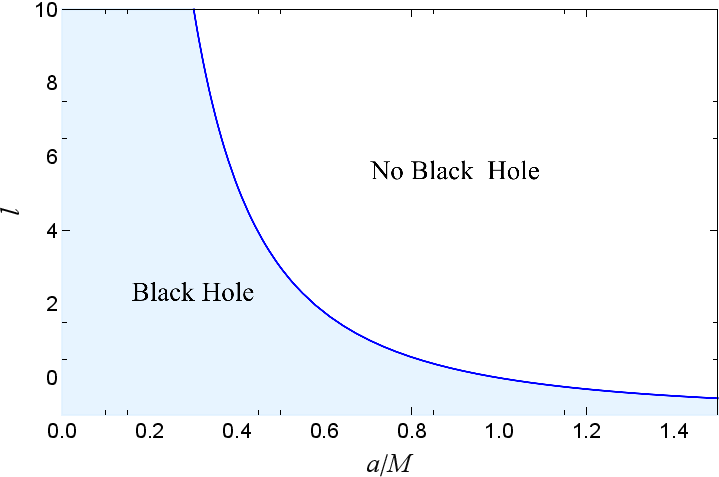}	
\caption{Parameter space ($a,l$). The blue region represents the allowed range of black holes}\label{parameter}	
\end{centering}
\end{figure}

We consider the bumblebee gravity model described by the following action~\citep{Casana:2017jkc,Ding:2019mal}:
\begin{equation}\label{eqn:100}
\begin{aligned}
    S=\int d^4 x \sqrt{-g} \Big[ & \frac{1}{16\pi} \left( R+\varrho B^\mu B^\nu R_{\mu\nu} \right) \\
    & -\frac{1}{4}B^{\mu\nu}B_{\mu\nu} - V(B^\mu) \Big] \, ,
\end{aligned}
\end{equation}
where $\varrho$ is a real coupling constant controlling the non-minimal gravity interaction to the bumblebee vector field $B^\mu$, $B_{\mu\nu}$ is the bumblebee field strength
\begin{equation}
    B_{\mu\nu}=\partial_{\mu}B_{\nu}-\partial_{\nu}B_{\mu} \, ,
\end{equation}
and $V(B^\mu)$ is a certain potential of the bumblebee vector field used to induce the violation of the Lorentz symmetry. The potential $V(B^\mu)$ has the form
\begin{equation}
    V = V (B^{\mu}B_{\mu}\pm b^2) \, ,
\end{equation}
where $b^2$ is a real positive constant. The potential must have a minimum at $B^{\mu}B_{\mu}\pm b^2 = 0$. The bumblebee field gets a non-vanishing vacuum expectation value $\langle B^\mu \rangle = b^\mu$, where $b^\mu$ is a vector field of constant norm: $b^\mu b_\mu = \mp b^2$ ($b^\mu$ can be either timelike or spacelike).

From the action in Eq.~(\ref{eqn:100}), we get the following field equations for the gravity sector:
\begin{align}
    R_{\mu\nu}-\frac{1}{2}Rg_{\mu\nu}= 8\pi T_{\mu\nu}^B \, , 
\end{align}
where the energy-momentum tensor of bumblebee field, $T_{\mu\nu}^B$, is given by~\citep{Casana:2017jkc}
\begin{equation}
\begin{aligned}\label{eqn:4}
    T_{\mu\nu}^B=&B_{\mu\alpha}{B^{\alpha}}_{\nu}-\frac{1}{4} g_{\mu\nu} B^{\alpha\beta}B_{\alpha\beta}-g_{\mu\nu}V+2B_\mu B_\nu V' \\
    & +\frac{\varrho}{8\pi} \Big[ \frac{1}{2} g_{\mu\nu} B^{\alpha}B^{\beta}R_{\alpha\beta}-B_{\mu}B^{\alpha}R_{\alpha\nu}-B_{\nu}B^{\alpha}R_{\alpha\mu} \\
    & +\frac{1}{2}\nabla_{\alpha}\nabla_{\mu}(B^{\alpha}B_{\nu})+\frac{1}{2}\nabla_{\alpha}\nabla_{\nu}(B^{\alpha}B_{\mu}) \\
    & -\frac{1}{2}\nabla^2(B_{\mu}B_{\nu})-\frac{1}{2}g_{\mu\nu}\nabla_{\alpha}\nabla_{\beta}(B^{\alpha}B^{\beta})\Big] \, ,
\end{aligned}
\end{equation}
and $V'$ is 
\begin{align}
    V'=\frac{\partial V(x)}{\partial x}\vert_{x=B^{\mu}B_{\mu}\pm b^2} \, .
\end{align}
The field equations of the bumblebee field are
\begin{align}
    \nabla^\mu B_{\mu\nu} = 2 V' B_\nu - \frac{\varrho}{8\pi} B^\mu R_{\mu\nu} , 
\end{align}
but in what follows, we will assume that the bumblebee field is frozen to its vacuum expectation value, namely $B^\mu = b^\mu$.

A spontaneous Lorentz symmetry breaking induces a vacuum solution when the bumblebee field $B_{\mu}$ remains frozen in its vacuum expectation value $b_{\mu}$. In this way, the bumblebee field is fixed to be 
\begin{equation}
    B_{\mu}=b_{\mu}
\end{equation}
and consequently, we have $V=V'=0$. Under such conditions, we have a Lorentz-violating spherically symmetric solution
\begin{equation}\label{metric1}
\begin{aligned}
 ds^2 = &-\left( 1-\frac{2M}{r} \right) dT^2  + (1+\ell)\left( 1-\frac{2M}{r} \right)^{-1} dr^2 \\
 ~~& + r^2 \theta^2 + r^2  \sin^2\theta d\phi^2,
\end{aligned}
\end{equation}
where we have defined the LSB parameter as $\ell=\varrho b^2$, which takes values in the range shown in Figure \ref{parameter}. The metric (\ref{metric1}) represents a purely radial Lorentz-violating solution outside a spherical body characterizing a modified black hole solution. By introducing a transformation, such that $t \to \sqrt{1+l}~T$, we observe that metric (\ref{metric1}) transforms into a Schwarzschild-like solution as:  
\begin{equation}\label{metric2}
\begin{aligned}
 ds^2 = &-(1+\ell)^{-1}\left( 1-\frac{2M}{r} \right) dt^2  + \frac{ dr^2}{ (1+\ell)^{-1}\left( 1-\frac{2M}{r} \right)} \\
 ~~& + r^2 \theta^2 + r^2  \sin^2\theta d\phi^2,
\end{aligned}
\end{equation}

\begin{figure}[t]
\begin{centering}
\includegraphics[scale=0.7]{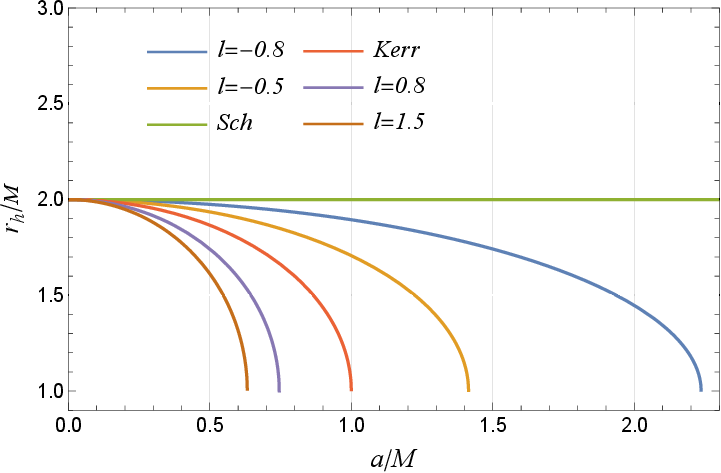}	
\caption{Variation in event horizon radii for the  RBHBG as a function of spin for different values of the LSB  parameter $\ell$.  We also provided Kerr and Schwarzschild event horizon variation for comparison.}\label{horizon1}	
\end{centering}
\end{figure}
The metric (\ref{metric1}) reduces to the Schwarzschild black hole without the LSB parameter, i.e., as $\ell \to 0$. While static black holes are theoretical and unlikely, spinning black holes are expected to be present in the universe and can be tested through astronomical observations. We then construct a Kerr-like metric as an axisymmetric generalization of the metric (\ref{metric2}) and validate it using EHT data. It is achieved with a modified Newman-Janis algorithm (NJA) \citep{Azreg-Ainou:2014pra,Azreg-Ainou:2014aqa}. The original Newman-Janis method \citep{Newman:1965tw} provides a groundbreaking technique to generate rotating spacetimes from a stationary, spherically symmetric initial metric without needing to solve field equations. By starting with a static and spherically symmetric black hole metric (\ref{metric2}) and applying the modified NJA \citep{Azreg-Ainou:2014pra,Azreg-Ainou:2014aqa}, we obtain the rotating spacetime given by

\begin{eqnarray}\label{metric3}
ds^2 &=& -\left(1-\frac{2M(r)r }{\Sigma}\right) dt^2+ \frac{\Sigma}{\Delta} dr^2 +\Sigma d\theta^2  \nonumber\\ && + \frac{4aM(r)r}{\Sigma} \sin^2\theta dtd\phi- \frac{\mathbb{A}\sin^2\theta~}{\Sigma} d\phi^2
\end{eqnarray}
where
\begin{eqnarray}
\Delta &=& r^2+a^2 -2 M(r) r~~~~~
\Sigma = r^2 +a^2\cos^2\theta,\nonumber \\
M(r) &=& \frac{M(1+\frac{r\ell}{2M})}{1+\ell}
~~~~~~~\mathbb{A} = (r^2+a^2)^2-a^2 \Delta \sin^2\theta. \nonumber \\
\end{eqnarray}
The black hole mass is denoted by $M$, the LSB parameter is $\ell$, and a specific spin parameter is $a$.  A non-vanishing value of $\ell$ results in a divergence from the Kerr solution, suggesting that the Lorentz symmetry is broken. For $\ell = 0$, we precisely recover spherical black hole \citep{Casana:2017jkc,Ding:2019mal}). We call the black holes represented by metric~(\ref{metric3}) as rotating black holes in Bumblebee gravity (RBHBG). 

The metric(\ref{metric3}) is singular at $\Sigma=0$ and at $\Delta=0$, with the singularity at $\Sigma=0$ is a ring-shaped physical singularity in the equatorial plane of the centre of a rotating black hole.  The radial coordinate of the event horizon may be determined using the equation $g^{rr}=\Delta=0$, much like for the Kerr spacetime. It comes out to be
\begin{align}
     r_{\rm h}=M + \sqrt{M^2-a^2(1+\ell)} \, ,
\end{align}
which requires
\begin{align}\label{eqn:11}
    |a| \leq \frac{M}{\sqrt{1+\ell}} \, .
\end{align}
If Eq.~(\ref{eqn:11}) is violated, the spacetime will feature a naked singularity without an event horizon. Our study will focus exclusively on the parameter space for black holes, excluding cases involving naked singularities. Using mass $M$ as the unit, Figure \ref{horizon1} shows the event horizon radius $r_{\rm h}$ as a function of spin $a$ for different values of the LSB parameter $\ell$. The numerical solutions match the Schwarzschild black hole when $a$ and $\ell$ are zero and the Kerr black hole when $\ell$ is zero. Notably, for $\ell < 0$, the maximum spin parameter can exceed $M$, while the horizon radius decreases with increasing $a$ for all $\ell$. In the spherical case ( $a =0$), the event horizon remains at $2M$ regardless of $\ell$.

\section{Black hole shadow}\label{Sec-3}
The black hole shadow is a dark region against the bright emissions of the accretion disk, defined by the photon sphere's boundary where the black hole's strong-gravitational field affects light paths. Photons follow null geodesics shaped by the black hole's mass, spin, or charge, creating a shadow surrounded by a bright photon ring \citep{Synge:1966,bardeen1973,Luminet:1979,Cunningham:1973}. This shadow reveals the spacetime structure around black holes and tests gravity theories in strong-field conditions. EHT observations are used to quantify black hole properties and assess theoretical predictions \citep{Vries:2000,Shen:2005cw,Amarilla:2010zq,Yumoto:2012kz,Amarilla:2013sj,Atamurotov:2013sca,Abdujabbarov:2016hnw,Abdujabbarov:2015xqa,Cunha:2018acu,Mizuno:2018lxz,Mishra:2019trb,Shaikh:2019fpu,Kumar:2020yem,Kumar:2018ple,Kramer:2004hd}.

\begingroup
\begin{table*}
	\caption{Equatorial circular prograde  ($r_{P}^{-}$) and retrograde  ($r_{P}^{+}$) photon orbit radii for RBHBG at two different values of the LSB  parameter i.e.,  $\ell=-0.2$ and $\ell=0.2$, compared with corresponding values for Kerr black boles $r_{K}^{\mp}$ at different values of spin.}\label{table1A}
	%%%%%%%%
	\begin{ruledtabular}
		\begin{tabular}{l c c c c c c}
			%%%%%%%%
			\multicolumn{1}{c}{}&
            \multicolumn{2}{c}{Kerr Black Hole }&
			\multicolumn{2}{c}{ $l=-0.2$}&
			\multicolumn{2}{c}{$l= 0.2$}\\
   \cmidrule{2-3} \cmidrule{4-5}  \cmidrule{6-7}
			%%%%%%%%%%%%%%%%%%%%%%%%%%%%%%%%%%%%%
$a/M$&$r_{K}^{+}/M $ & $r_{K}^{-}/M $ & $r_{P}^{+}/M $ & $r_{P}^{-}/M $ &$r_{P}^{+}/M $ & $r_{P}^{-}/M $ \\
\hline
0. & 3. & 3. & 3. & 3. & 3. & 3. \\
0.1 & 3.11335 & 2.88219 & 3.10157 & 2.89486 & 3.12396 & 2.87069 \\
0.2 & 3.22281 & 2.75919 & 3.2 & 2.78564 & 3.24331 & 2.73505 \\
0.3 & 3.32885 & 2.63003 & 3.29562 & 2.67167 & 3.35864 & 2.59173 \\
0.4 & 3.43184 & 2.49336 & 3.3887 & 2.55209 & 3.47042 & 2.43884 \\
0.5 & 3.53209 & 2.3473 & 3.4795 & 2.42572 & 3.57904 & 2.27349 \\
0.6 & 3.62985 & 2.18891 & 3.5682 & 2.29087 & 3.6848 & 2.09092 \\
0.7 & 3.72535 & 2.01333 & 3.65498 & 2.14502 & 3.78798 & 1.88212 \\
0.8 & 3.81876 & 1.81109 & 3.74 & 1.984 & 3.8888 & 1.6251 \\
0.9 & 3.91027 & 1.55785 & 3.82337 & 1.8 & 3.98745 & 1.2 \\
		\end{tabular}
	\end{ruledtabular}
\end{table*}
\endgroup
To find the null geodesics of photons in the RBHBG spacetime, we use the Hamilton-Jacobi equation \citep{Carter:1968rr,Chandrasekhar:1985kt}. The metric (\ref{metric3}) is invariant under time translation and rotation, leading to conserved quantities such as energy $\mathcal{E}=-p_t$ and axial angular momentum $\mathcal{L}=p_{\phi}$. Therefore, we can determine the first-order differential equations of motion from the four integrals of motion: the Lagrangian, energy $\mathcal{E}$, axial angular momentum $\mathcal{L}$, and the Carter constant \citep{Carter:1968rr,Chandrasekhar:1985kt}.
\begin{eqnarray}
\Sigma \frac{dt}{d\tau}&=&\frac{r^2+a^2}{\Delta}\left({\cal E}(r^2+a^2)-a{\cal L}\right)  \nonumber\\ &&-a(a{\cal E}\sin^2\theta-{\mathcal {L}}),~ \label{tuch}\\
\Sigma \frac{d\phi}{d\tau}&=&\frac{a}{\Delta}\left({\cal E}(r^2+a^2)-a{\cal L}\right)-\left(a{\cal E}-\frac{{\cal L}}{\sin^2\theta}\right),\label{phiuch}\\
\Sigma \frac{dr}{d\tau}&=&\pm\sqrt{\mathcal{R}(r)}\ ,\label{r}\\
\Sigma \frac{d\theta}{d\tau}&=&\pm\sqrt{\Theta(\theta)}\ ,\label{th}
\end{eqnarray}

where  $\mathcal{R}(r)$ and $\Theta(\theta)$, respectively, pertain to the following radial and polar motion effective potentials: 
\begin{eqnarray}\label{06}
\mathcal{R}(r)&=&\left[(r^2+a^2){\cal E}-a{\cal L}\right]^2-\Delta[{\cal K}+(a{\cal E}-{\cal L})^2]\label{rpot},\quad \\ 
\Theta(\theta)&=&{\cal K}-\left[\frac{{\cal L}^2}{\sin^2\theta}-a^2 {\cal E}^2\right]\cos^2\theta,\label{theta0}
\end{eqnarray}
The Carter constant $\mathcal{K}$ and the separability constant $\mathcal{Q}$ are related by $\mathcal{Q} = \mathcal{K} + (a\mathcal{E} - \mathcal{L})^2$ \citep{Carter:1968rr,Chandrasekhar:1985kt}, reflecting the isometry of Equation (\ref{metric3}) along the second-order Killing tensor field. While $\mathcal{E}$ and $\mathcal{L}$ are linked to spacetime symmetries, the Carter constant $\mathcal{K}$ is not. The constants $\mathcal{Q}$ and $\mathcal{L}$ govern the $\phi$ and $\theta$ motions, respectively. When $\mathcal{Q}=0$, photons are restricted to an equatorial plane. Unlike Schwarzschild black holes, which have planar null circular orbits due to spherical symmetry, rotating black holes exhibit non-planar orbits due to frame dragging.

The black hole shadow silhouette  is outlined by the unstable spherical photon orbits, which can be determined by solving $\dot{r}=\ddot{r}=0$ from Eqs.~(\ref{r}) and (\ref{06}). The radii $r_p$ of the  photon orbits is positive root of the following equations:
\begin{equation}
\mathcal{R}|_{r=r_p}=\left.\frac{\partial \mathcal{R}}{\partial r}\right|_{r=r_p}=0,\,\, \text{and}\,\, \left.\frac{\partial^2 \mathcal{R}}{\partial r^2}\right|_{r=r_p}> 0.\label{vr} 
\end{equation}
To proceed further, following Chandershaker (\citep{Chandrasekhar:1985kt}), we can introduce the dimensionless parameters  
$\xi\equiv \mathcal{L}/\mathcal{E},\qquad \eta\equiv \mathcal{K}/\mathcal{E}^2$ to reduce the degree of freedom of photons geodesics to one.  Solving Eq.~(\ref{vr}) for Eq.~(\ref{rpot}) results in the critical impact parameters as follows: \citep{Chandrasekhar:1985kt}
\begin{eqnarray}\label{impactparameter}
\xi_c &=& \frac{a^2(r+ M+2rl )+r^2(r - 3 M )}{a (M- r)}\nonumber\\
\eta_c &=& \frac{ r^3 \left[4 a^2(1+l)  M - r(r - 3 M )^2\right]}{a^2 (M- r)^2}~~~~~
\end{eqnarray}
Here, $ '$ denotes the derivative with respect to the radial coordinate. In the limit $l \to 0$, Eq.~(\ref{impactparameter}) reduces to the critical impact parameter for the Kerr black hole \citep{Chandrasekhar:1985kt}. Light rays from a strong source follow three trajectories: (i) capture orbit, (ii) scatter orbit, and (iii) unstable orbit. The black hole's shadow is formed by light beams falling into the black hole, with unstable photon orbits marking the boundary between capture and scatter regions. At the equatorial plane, there are two types of circular photon orbits: prograde, moving in the same direction as the black hole's rotation, and retrograde, moving in the opposite direction. Due to the Lense-Thirring effect, prograde orbits are closer to the black hole than retrograde orbits, as the rotation of spacetime reduces the effective gravitational force in the direction of the spin. The radii of the prograde ($r_{p}^{-}$) and retrograde ($r_{p}^{+}$) orbits are obtained as roots of $\eta_c = 0$, leading to the following equation
\begin{equation}\label{eq23}
r^3-6 M r^2+9 M^2 r-4  M a^2  (1+l)=0.
\end{equation}
The two primary solutions of the Eq. (\ref{eq23}),  which represent the radii of the prograde $r_{p}^{-}$ and retrograde photon orbits $r_{p}^{+}$, respectively, are given by
\begin{eqnarray}
    r_{p}^{\mp}=2M\left[ 1 + \cos\left( \frac{2}{3}\arccos\left(\mp\frac{a\sqrt{1+l}}{M}  \right)\right)\right].
\end{eqnarray}

\begin{figure*}
    \begin{tabular}{p{9cm} p{9cm}}
     \includegraphics[scale=0.7]{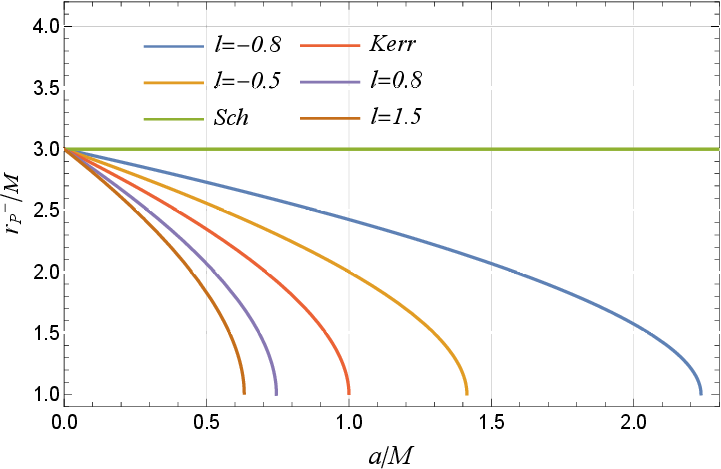}&
    \includegraphics[scale=0.7]{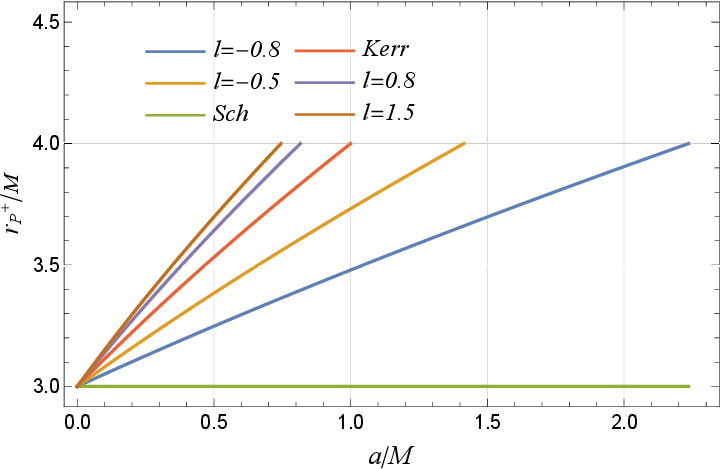}
    \end{tabular}
    \caption{Variation in the equatorial prograde (\textit{Left}) and retrograde photon sphere radii (\textit{Right}) for the RBHBG as a function of spin for different values of the LSB  parameter $\ell$.  We include also, for comparison, the variation in the Kerr  and Schwarzschild radii in both cases.}\label{photonOrbit}
\end{figure*}
\begin{figure*}
    \begin{tabular}{p{8cm} p{8cm}}
   \includegraphics[scale=0.72]{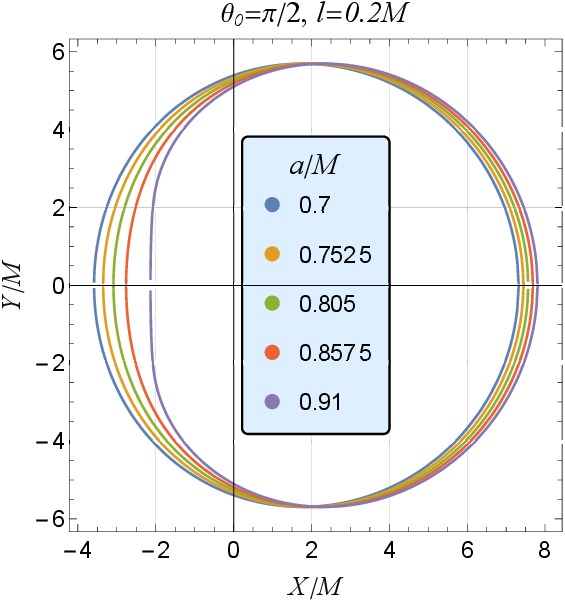}&
    \includegraphics[scale=0.7]{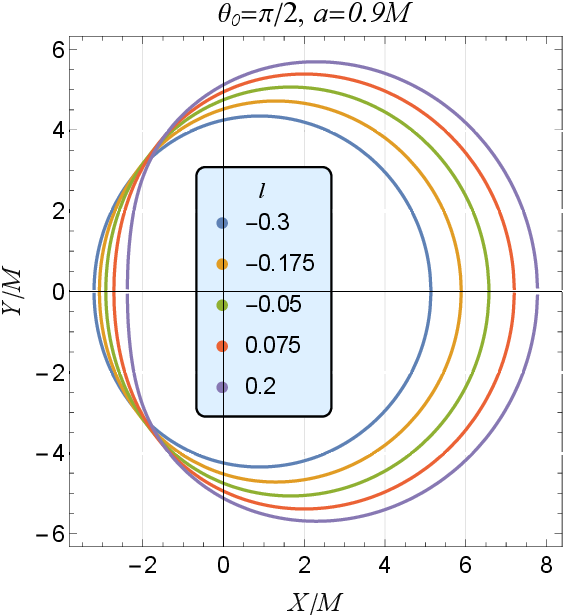}\\
     \includegraphics[scale=0.72]{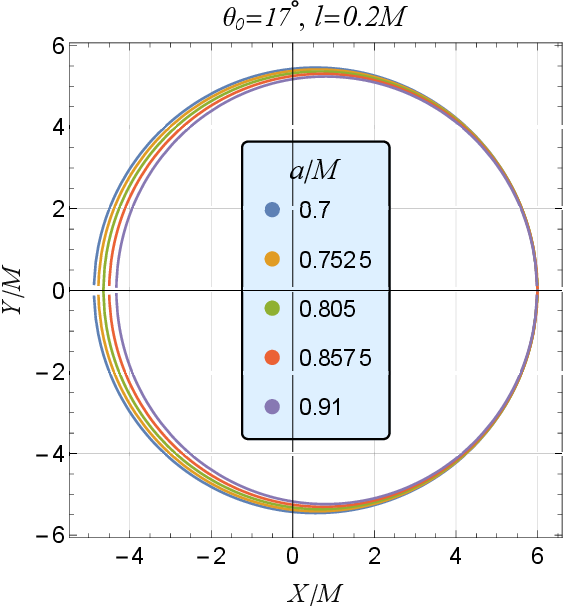}&
    \includegraphics[scale=0.7]{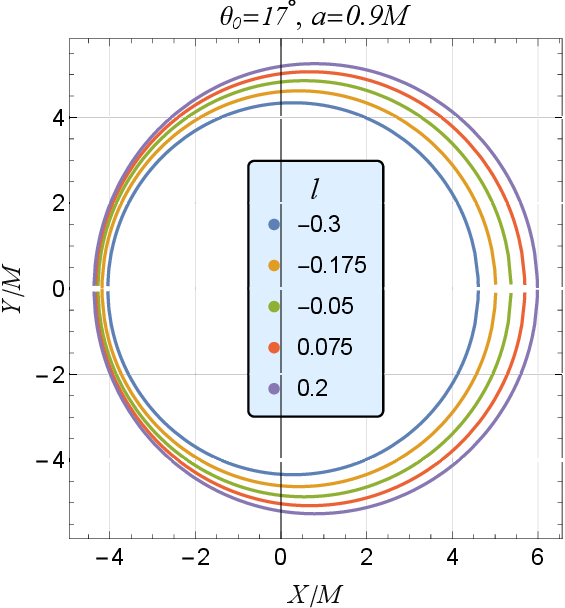}
    \end{tabular}
    \caption{Shadow silhouette of the RBHBG for $l=0.5$ with varying $a$  (\textit{left}) and for $a=0.95$ with varying $l$ (\textit{right}) as seen from the equatorial plane, i.e., inclination angle $\theta_o=\pi/2$ and $\theta_o=17\degree$.}\label{fig2}
\end{figure*}
The photon orbits $r_{p}^{-}$ and $r_{p}^{+}$ vary inversely with the parameter $a$, where $r_{p}^{-}$ decreases and $r_{p}^{+}$ increases as $a$ changes. In Bumblebee gravity, the LSB parameter $\ell$ introduces deviations from GR, affecting the black hole metrics. Table \ref{table1} shows the photon sphere radii $r_{p}^{-}$ and $r_{p}^{+}$, highlighting how Lorentz violation influences the photon sphere structure. As $\ell$ increases, $r_{p}^{+}$ increases while $r_{p}^{-}$ decreases (cf. Figure~\ref{photonOrbit} and Table \ref{table1}). Compared to Kerr black holes at a constant $a$, $r_{p}^{+}$ is lower for RBHBG when $\ell < 0$ and higher otherwise, while $r_{p}^{-}$ is greater for RBHBG when $\ell < 0$ and smaller otherwise (cf. Table \ref{table1}). Photon orbit radii in RBHBG, like Kerr black holes, depend explicitly on spin and, for all $\ell$, range from $M \leq r_{p}^{-} \leq 3M$ and $3M \leq r_{p}^{+} \leq 4M$ (cf. Figure~\ref{photonOrbit}). Additionally, non-planar photon orbit radii for RBHBG fall within $r_{p}^{-} \leq r_{p} \leq r_{p}^{+}$.

\subsection{Shadow Silhouette}
Plotting the black hole silhouette involves visualizing the apparent boundary of a black hole as seen from a distance, often referred to as the photon sphere. This shadow is defined by photons on the edge of being captured by the black hole's gravitational pull but manages to escape. The shape of the silhouette is affected by the black hole's spin (or angular momentum) and the angle of observation or inclination.   Using the tetrad components of the four momentum $p^{(\mu)}$ and geodesic Eqs.~(\ref{tuch}),(\ref{phiuch}), (\ref{r}) and (\ref{th}), a relationship between the observer's celestial coordinates, $X$ and $Y$, and two constants, $\xi_c$ and $\eta_c$ is deduced as follows
\begin{eqnarray}
&&X= -r_o\frac{p^{(\phi)}}{p^{(t)}} = -\left. r_o\frac{\xi_c}{\sqrt{g_{\phi\phi}}(\zeta-\gamma\xi_c)}\right|_{(r_o,\theta_o)},\nonumber\\
&&Y = r_o\frac{p^{(\theta)}}{p^{(t)}} =\pm\left. r_o\frac{\sqrt{\Theta(\theta)}
}{\sqrt{g_{\theta\theta}}(\zeta-\gamma\xi_c)}\right|_{(r_o,\theta_o)},~\label{Celestial}
\end{eqnarray} 
where 
\begin{eqnarray}
\zeta=\sqrt{\frac{g_{\phi\phi}}{g_{t\phi}^2-g_{tt}g_{\phi\phi}}},\qquad \gamma=-\frac{g_{t\phi}}{g_{\phi\phi}}\zeta.
\end{eqnarray}

The coordinates $X$ and $Y$ in Eq.~(\ref{Celestial}) represent the apparent displacement along the perpendicular and parallel axes to the projected axis of the black hole symmetry, respectively.  Therefore, an individual can visually perceive the stereographic projection of the black hole's shadow, which is determined by the celestial coordinates specified by Bardeen \citep{bardeen1973}, at an infinite radial distance ($r_o\to\infty$) and an inclination angle of $\theta_0$ as
\begin{equation}\label{pt}
X=-\xi_c\csc\theta_o,\qquad Y=\pm\sqrt{\eta_c+a^2\cos^2\theta_o-\xi_c^2\cot^2\theta_o}.
\end{equation} 
For an observer in the equatorial plane ($\theta_0=\pi/2$), Eq.~(\ref{pt}), reduces to
\begin{equation}\label{eq28}
    \{X,Y\}=\{-\xi_c,\pm\sqrt{\eta_c}\},
\end{equation}
and whereas for $\ell = 0$, Eq.~\ref{eq28}, reduces to
\begin{eqnarray}\label{eq29}
X_{K} &=& \frac{a^2(r_{p}+ M )+r_{p}^2(r - 3 M )}{a (M- r_{p})}\nonumber\\
Y_{K} &=& \frac{ r_{p}^{3/2} \left[4 a^2  M - r_{p}(r_{p} - 3 M )^2\right]^{1/2}}{a (M- r_{p})}~~~~~
\end{eqnarray}
which is exactly same as obtained for the Kerr black hole \citep{Hioki:2009na}. The parametric plots of Eq. (\ref{pt}) provide a Schwarzschild-like shadow for $a=0$, with the contour given by $X^2+Y^2=27(1+\ell)M^2$. The shadow of nonrotating black holes in Bumblebee gravity is larger than that of Schwarzschild black holes and increases with $\ell$. Figure~\ref{fig2} illustrates the RBHBG spacetime shadow silhouette for various parameter values. As $\ell$ increases, the shadow size grows and becomes distorted for a fixed spin $a$. The shadow shifts horizontally along the $X$-axis with increasing inclination angle $\theta_0$ and spin $a$. Unlike Kerr black holes, where the shadow center is always positive, in RBHBG, the center can be negative for small values of $a$
\begin{figure}
\begin{center}
	\includegraphics[scale=0.85]{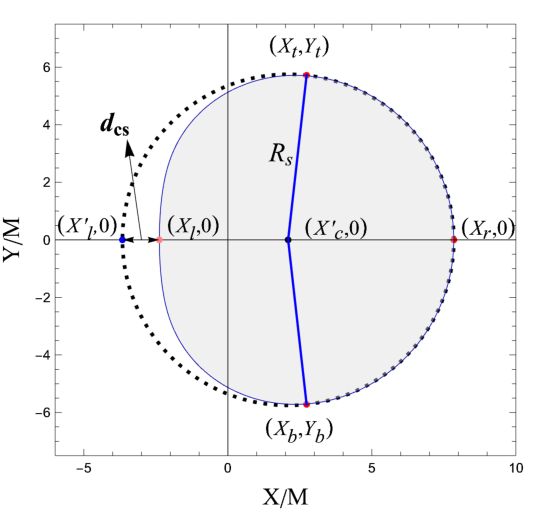}
	\caption{Figure presenting  the shadow observables, radius $R_s$ and the distortion parameter $\delta_s = d_{cs}/R_s$, for the apparent shape of the black hole's shadow within the context of Bumblebee Gravity. The chosen parameters ($a=0.9M$, $l=0.2$, $\theta_0=\pi/2$) correspond to a rapidly rotating black hole within the Bumblebee Gravity framework. }\label{dis}
\end{center}
\end{figure}
\section{Parameter Estimation of black holes}\label{Sec-4}
The motivation for parameter estimation of black holes is rooted in testing and constraining fundamental theories of gravity, including the "no-hair theorem," which posits that black holes are fully described by just three parameters: mass, spin, and charge \citep{Werner:1967ab,Werner:1968ab,Carter:1971zc,Misner:1973prb}. We can challenge or refine the no-hair theorem by estimating these parameters and exploring deviations, such as Lorentz symmetry breaking or other modifications beyond GR \citep{Werner:1967ab}. Observational data from instruments like the Event Horizon Telescope (EHT) \citep{EventHorizonTelescope:2019dse} offer a unique opportunity to investigate black holes in extreme conditions, validate theoretical models, and potentially uncover new physics that challenges conventional assumptions. Precise parameter estimation thus plays a crucial role in advancing our understanding of black hole behaviour and the nature of spacetime.

The black hole shadow (Figure~\ref{fig2}) is a critical observable that reveals the black hole's properties and spacetime geometry. Scientists can test GR, explore alternative gravity theories, and constrain deviation parameters by measuring its shape and size. As shown earlier, the black hole's rotation and the LSB parameter $\ell$ introduce asymmetry in the shadow, with higher spin and $\ell$ values causing increased distortion. In Bumblebee gravity, a crucial result is the ability to estimate the parameter $\ell$ through observational data.

First, we employ the method proposed by Hioki and Maeda \citep{Hioki:2009na} for parameter estimation using shadow observables, specifically the radius $R_s$ and distortion $\delta_s$, allowing for precise determination of black hole properties through deviations in shadow size and shape. Building on this, we apply the Kumar and Ghosh method \citep{Kumar:2018ple}, which focuses on the shadow area $A$ and oblateness $D$. By incorporating these observables, we improve our estimates of parameters like spin $a$ and the LSB parameter $\ell$.  Black hole parameters can be back-estimated using prior knowledge from observing these observables. Our theoretical study seeks to regulate black hole parameters like LSB  parameter $\ell$. However, errors in mass and distance measurements have been accounted for in EHT results.  Assuming priors on mass and distance, we find that for M87* the mass $M=\left(6.5\pm0.7\right)\times{10}^9M_\odot$, and its distance is $d=16.8Mpc$ \citep{EventHorizonTelescope:2019ggy} and that of  SgrA* is $M=4_{-0.6}^{+1.1}\times{10}^6M_\odot$, and its distance is $d=7.97 kpc$ \citep{Chen:2019tdb}.  

We start with defining the two observables radius $R_s$ and distortion $\delta_s$, to characterize the black hole shadow silhouette as follows \citep{Hioki:2009na}:
\begin{eqnarray}
R_s &=& \frac{(X_t-X_r)^2+Y_t^2}{2|X_r-X_t|},\nonumber\\
\delta_s &=& \frac{|X'_l-X_l|}{R_s}.
\end{eqnarray}
A reference perfect circle with a center $(X'_c, 0)$ that coincides with the shadow silhouette at three points, $(X_t,Y_t)$,  $(X_b,Y_b)$,  $(X_r,0)$,  to approximate the shape of the black hole shadow is drawn as shown in Figure ~\ref{dis}. The radius  $R_s$  of the shadow is defined  by the radius of this reference circle. Further, the points $(X_l,0)$,  $(X'_l,0)$, represent the intersections of the shadow silhouette and reference circle with the horizontal axis $(Y=0)$, respectively, such that $d_{cs}=|X'_l-X_l|$ determines the potential dent on the black hole shadow in the direction perpendicular to the black hole rotational axis. 
\begin{figure*}
    \begin{tabular}{p{9cm} p{9cm}}
     \includegraphics[scale=0.65]{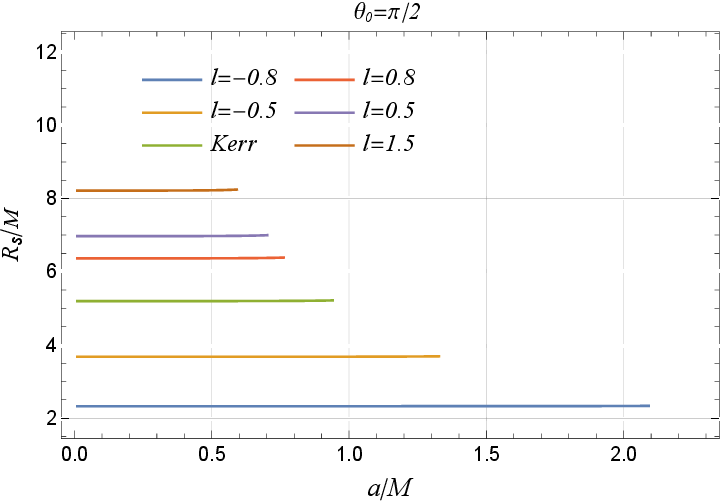}&
     \includegraphics[scale=0.65]{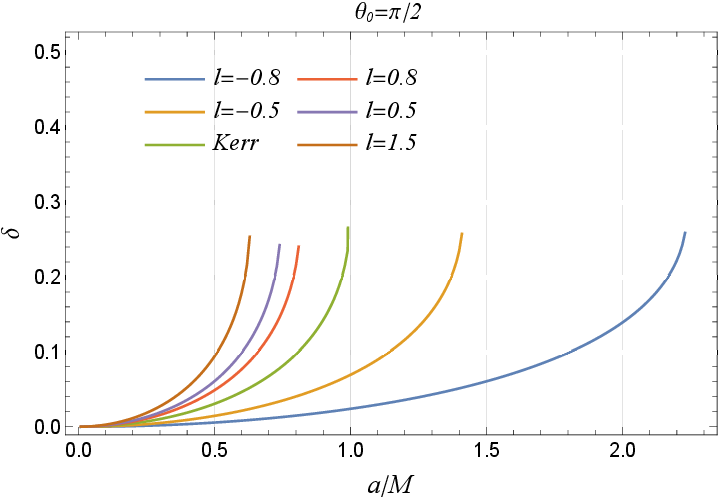}\\
     \includegraphics[scale=0.65]{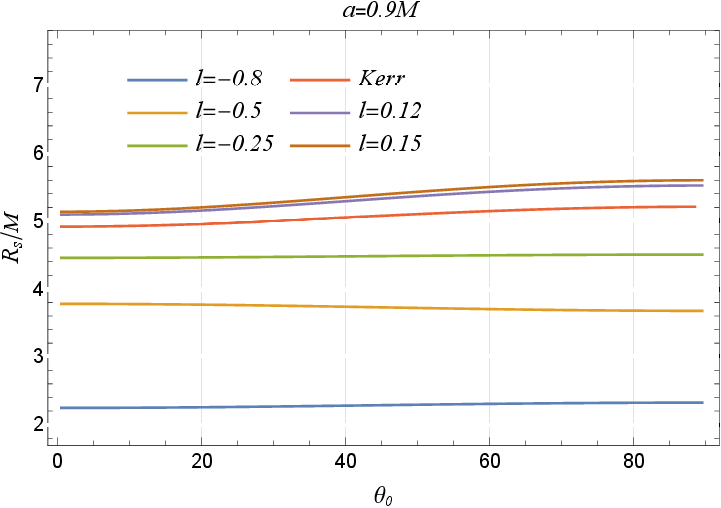}&
      \includegraphics[scale=0.65]{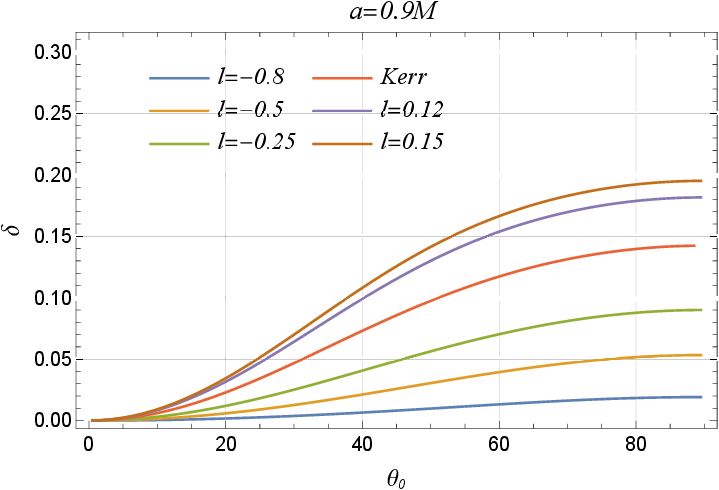}\\
     \includegraphics[scale=0.65]{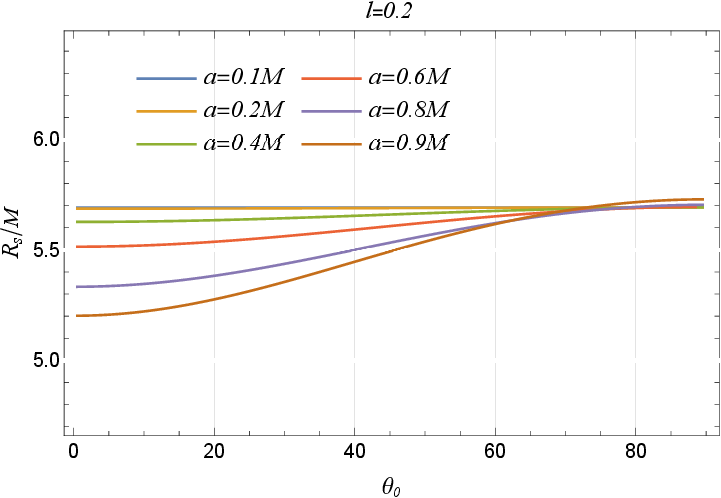}&
      \includegraphics[scale=0.65]{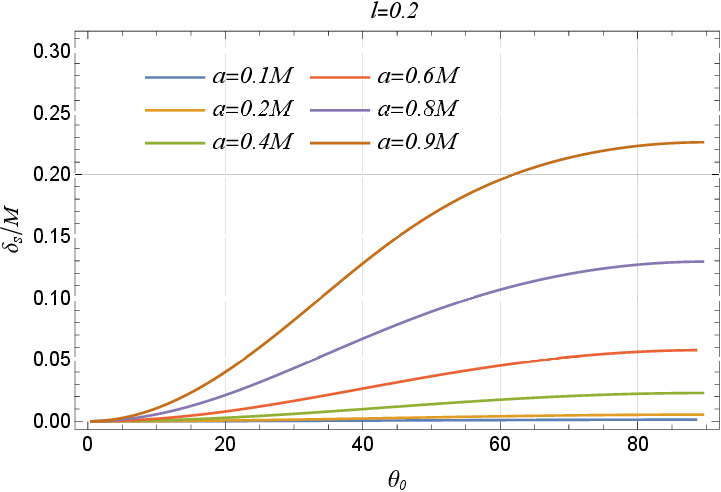}
    \end{tabular}
    \caption{The plots illustrate the impact of the LSB parameter, spin, and inclination angle $\theta_0$ on two crucial black hole shadow observables: the radius of the reference perfect circle that coincides with the black hole shadow and the distortion in the black hole shadow. The plots are divided into three panels, each focusing on varying two parameters while keeping the third one fixed.  The inclination angle is fixed at $\theta_0=\pi/2$, while the LSB parameter and spin are varied  (\textit{Top panel}).  The spin parameter  is fixed at $a=0.9M$, while the LSB parameter and inclination angles are varied (\textit{Middle panel}). The LSB parameter is fixed at $\ell=0.2$, while the spin and  $\theta_0$ are varied  (\textit{Bottam panel}).  The corresponding changes in the radius of the reference perfect circle (\textit{Left}) and distortion in the black hole shadow  (\textit{Right}) are highlighted. }\label{fig6A}
\end{figure*}
The LSB parameter $\ell$ consistently increases the black hole shadow radius, regardless of spin parameter $a$ or inclination angle $\theta_0$, significantly altering black hole spacetime (cf. Figure~\ref{fig6A}). In some cases, $\ell$ and  $\theta_0$ can overshadow the effects of spin on the shadow radius. While spin has minimal impact at fixed inclination angles, it becomes less relevant at high inclinations with spin-independent radius. At high spins, the shadow's properties are dominated by spin, with $ \theta_0$ effects being more noticeable at lower spins (cf. Figure~\ref{fig6A}). The LSB parameter $\ell$ primarily affects shadow distortion $\delta_s$ at high inclination angles and spins, with its impact being less noticeable at lower inclinations. At fixed inclinations, distortion is mainly influenced by spin, increasing exponentially, while both $a$ and $\ell$ enhance distortion at higher inclinations, though their effects are subdued at lower angles (cf. Figure~\ref{fig6A}).
\begin{figure*}
    \begin{tabular}{p{9cm} p{9cm}}
    \includegraphics[scale=0.65]{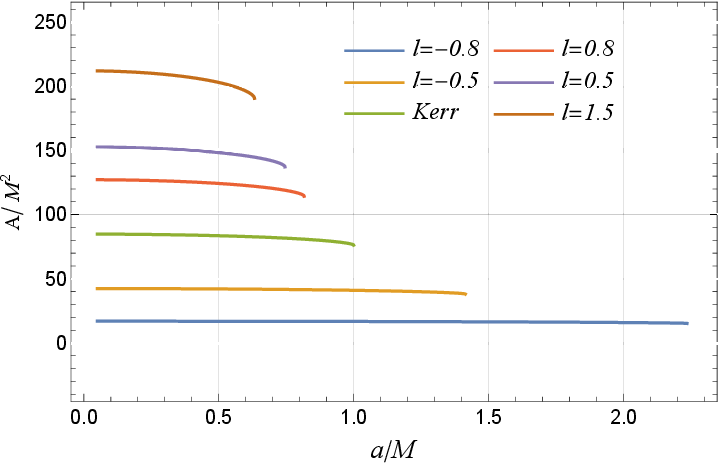}&
     \includegraphics[scale=0.65]{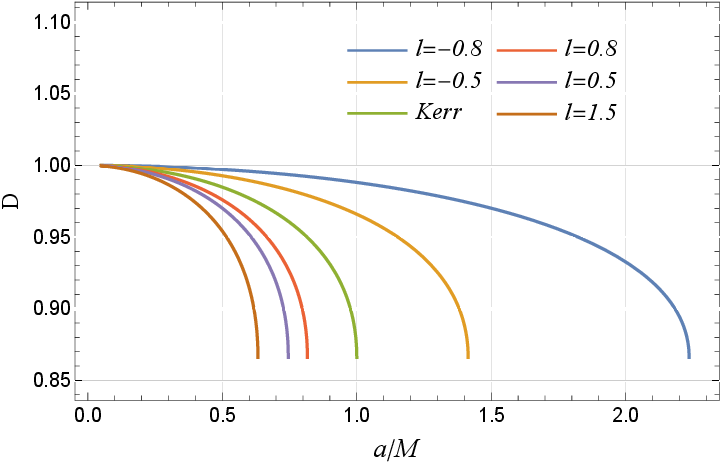}\\
     \includegraphics[scale=0.65]{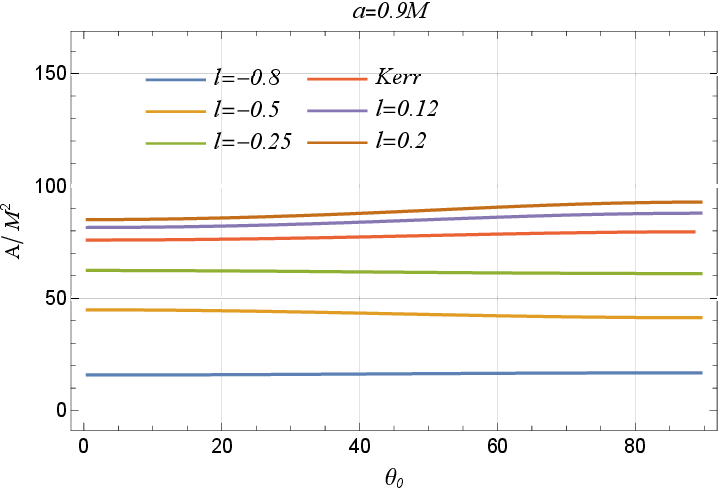}&
    \includegraphics[scale=0.65]{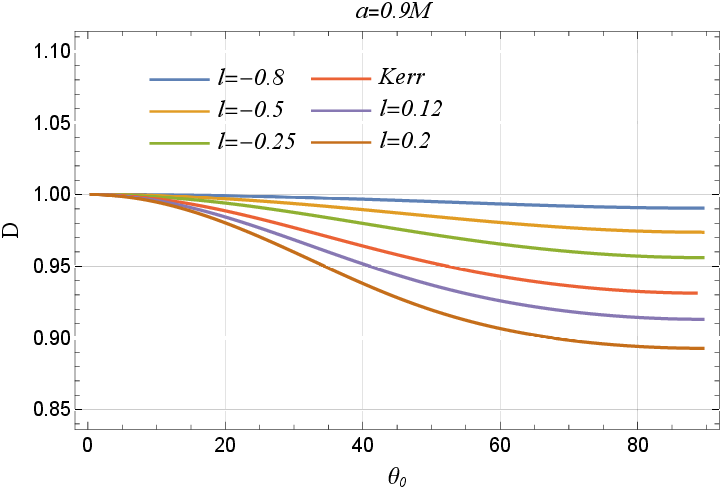}\\
     \includegraphics[scale=0.65]{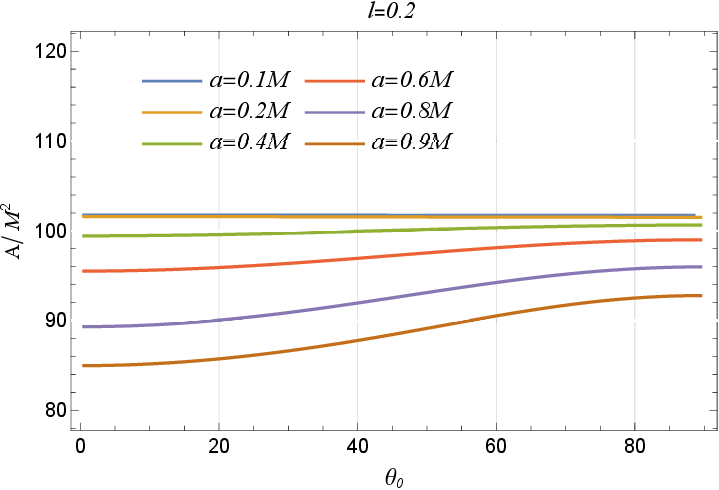}&
    \includegraphics[scale=0.65]{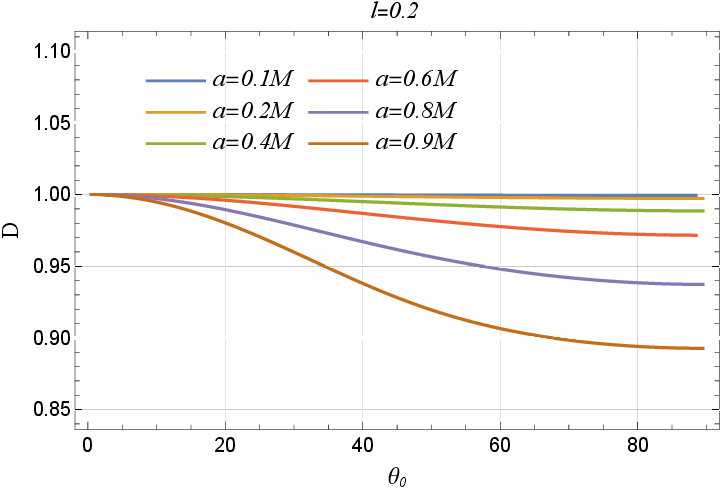}
    \end{tabular}
    \caption{Set of plots that illustrates the impact of the LSB parameter, spin, and inclination angle on two important shadow observables of black holes: the average area of the black hole shadow and the oblateness of the black hole shadow. The plots are divided into three panels, each focusing on varying two parameters while keeping the third one fixed.  The $ \theta_0$  is fixed at   $\theta_0=\pi/2$, while the LSB parameter and spin are varied (\textit{Top Panel}). The spin parameter  is fixed at $a=0.9M$, while the LSB parameter and $ \theta_0$ s are varied  (\textit{Middle panel}).  The LSB parameter is fixed at $\ell=0.2$, while the spin and $ \theta_0$  are varied (\textit{Bottam panel}).  The corresponding changes in the the average area of the black hole shadow  (\textit{Left}) and the Oblateness (\textit{Right}) are highlighted.}\label{fig6B}
\end{figure*}
The black hole parameters are determined by combining contour plots of the radius and distortion observables, establishing a direct correlation with parameters ($a$, $\ell$) and values ($R_s$, $\delta_s$). With fixed black hole mass $M$ and $ \theta_0$ , and measured $R_s$ and $\delta_s$, the spin parameter $a$ and LSB parameter $\ell$ can be accurately calculated using Figure \ref{contour1} and Table \ref{table1}. For instance, for a black hole viewed at an inclination of $\theta_{0}=\pi/2$ with mass $M$ and observed $R_s=4.8M$ and $\delta_s=0.15$, we can infer $a=0.982M$ and $\ell=-0.1487$. This method provides a reliable estimate of $a$ and $\ell$.
\begin{figure}
\begin{center}
	\includegraphics[scale=0.65]{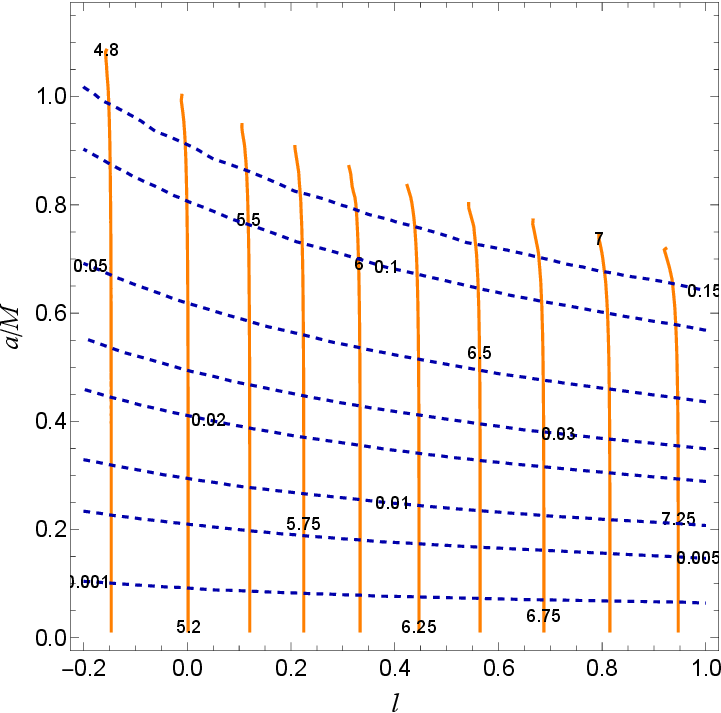}
	\caption{This contour plot displays the shadow observables $R_s/M$ (red solid lines), representing the normalized shadow radius, and $\delta_s$ (blue dashed lines), which quantifies the shadow's distortion. The intersection of these contours at a unique point in the parameter space ($a$, $l$) pinpoints the specific values of the black hole's spin and LSB parameters. }\label{contour1}
\end{center}
\end{figure}
\begin{table}
\caption{Estimated Values of Parameters ($a/M$, $l$) from Contour Plots of Shadow Observables $R_s$ and $\delta_s$.}\label{table1}
	\centering	
	\begin{tabular}
	{ |p{1.6cm}|p{1.6cm}|p{1.6cm}|p{1.6cm}| }
		\hline
		$R_s/M$ &  $\delta_s$ & $a/M$ & $l$  \\
		\hline\hline
 5.24 & 0.01 & 0.2866 & 0.0175 \\
 \hline
 5.24 & 0.14 & 0.8851 & 0.0126 \\
 \hline
 5.32 & 0.02 & 0.395 & 0.0485 \\
 \hline
 5.32 & 0.18 & 0.923 & 0.0404 \\
 \hline
 5.44 & 0.05 & 0.5914 & 0.0936 \\
 \hline
 5.44 & 0.18 & 0.9053 & 0.086 \\
 \hline
 5.52 & 0.10 & 0.756 & 0.1224 \\
 \hline
 5.52 & 0.18 & 0.8939 & 0.1167 \\
 \hline
 5.64 & 0.08 & 0.6872 & 0.1694 \\
 \hline
 5.66 & 0.005 & 0.1882 & 0.1789 \\
 \hline
 5.66 & 0.10 & 0.7415 & 0.1761 \\
 \hline
\end{tabular}
\end{table}

\begin{figure}
\begin{center}
	\includegraphics[scale=0.65]{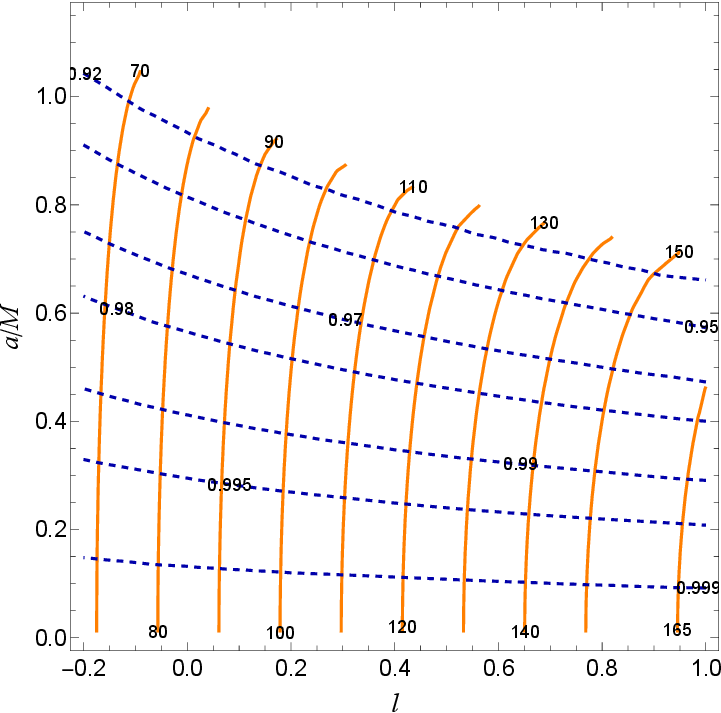}
	\caption{The contour plot illustrates the shadow observables $A/M^2$ (red solid lines), representing the dimensionless area of the black hole's shadow, and $D$ (blue dashed lines), which measures the distortion of the shadow. The intersection of these contours at a unique point in the parameter space ($a$, $l$) identifies the specific values of the black hole's spin and LSB parameters.  }\label{contour2}
\end{center}
\end{figure}

\begin{table}
\caption{Estimated Values of Parameters ($a$, $l$) from Contour Plots of Shadow Observables $A/M^2$ and $D$.}\label{table2}
	\centering	
	\begin{tabular}
	{ |p{1.6cm}|p{1.6cm}|p{1.6cm}|p{1.6cm}| }
		\hline
		$A/M^2$ &  $D$ & $a/M$ & $l$  \\
		\hline\hline
 
 82 & 0.94 & 0.8503 & 0.0229 \\
 \hline
 84 & 0.96 & 0.7367 & 0.0295 \\
 \hline
 86 & 0.98 & 0.5498 & 0.0343 \\
 \hline
 88 & 0.94 & 0.8206 & 0.096 \\
 \hline
 90 & 0.995 & 0.2823 & 0.066 \\
 \hline
 90 & 0.91 & 0.8896 & 0.1447 \\
 \hline
 92 & 0.91 & 0.8787 & 0.1683 \\
 \hline
 92 & 0.99 & 0.3893 & 0.0935 \\
 \hline
 94 & 0.95 & 0.7539 & 0.1564 \\
 \hline
 96 & 0.97 & 0.6181 & 0.1596 \\
 \hline
 99 & 0.99 & 0.376 & 0.1725 \\
 \hline
\end{tabular}
\end{table}

We can also employ the black hole shadow observables, the shadow oblateness ($D$), and the area ($A$) enclosed by a black hole shadow.  The observables  are defined as follows: 
\begin{eqnarray}
    A &=& 2\int_{r_p^{-}}^{r_p^+}\left( Y(r_p) \frac{dX(r_p)}{dr_p}\right)dr_p,\\
    D &=& \frac{X_r-X_l}{Y_t-Y_b}.
\end{eqnarray}\label{Area1}
The shadow silhouette's edges are denoted by the subscripts $r$, $l$, $t$, and $b$ for the right, left, top, and bottom, respectively. For a spherically symmetric black hole, $D=1$, while for a Kerr black hole, $\sqrt{3}/2 \leq D < 1$ \citep{Tsupko:2017rdo}. The shadow area depends on the LSB parameter $\ell$, spin, and $ \theta_0$ . As $\ell$ increases, the shadow area grows if the spin and $ \theta_0$  are fixed, indicating a direct impact of $\ell$ on the shadow size. At higher spin values, the shadow area remains stable across different inclination angles, but at lower inclinations, the shadow area varies significantly with spin when $\ell$ is constant. This effect diminishes at higher inclinations, suggesting that the inclination angle's impact on the shadow area is more pronounced at lower angles (cf. Figure~\ref{fig6B}).
\begingroup
\begin{table*}
	\caption{The table presents a comprehensive estimation of the event horizon size for a given spin parameter of $a=0.9M$ and different values of $\ell$.  Additional characteristics include the measurements of the radius of both prograde and retrograde circular photon orbits ($r^{\mp}_P$), as well as the area of the event horizon ($A_{H}$). In addition, the table displays the average size of the black hole's shadow ($A$). The ratio between the area of the event horizon and the average area of the black hole shadow is given, providing information about the geometric and observational properties of the black hole Sgr A*. 	}\label{table1B}
	%%%%%%%%
	\begin{ruledtabular}
		\begin{tabular}{l c c c c c c}

$l$ &$r_{h}/M $ & $r_{P}^{+}/M $ & $r_{P}^{-}/M $ & $A_{H}*(10^{22})(m^2) $ &$A*(10^{22})(m^2)$ & $A/A_{H} $ \\
			\hline
 -0.8 & 1.91542 & 3.43437 & 2.48985 & 4.63076 & 6.75176 & 1.45802 \\
 -0.7 & 1.87006 & 3.52511 & 2.35796 & 4.41401 & 10.0762 & 2.28277 \\
 -0.6 & 1.82219 & 3.6     & 2.23923 & 4.19094 & 13.3626 & 3.18844 \\
 -0.5 & 1.77136 & 3.66486 & 2.12736 & 3.96039 & 16.6074 & 4.19338 \\
 -0.4 & 1.71694 & 3.72264 & 2.01867 & 3.72076 & 19.8062 & 5.32315 \\
 -0.3 & 1.65803 & 3.7751  & 1.91043 & 3.46982 & 22.9529 & 6.61503 \\
 -0.2 & 1.5933 & 3.82337  & 1.8     & 3.20417 & 26.0395 & 8.12675 \\
 -0.1 & 1.52058 & 3.86824 & 1.68416 & 2.91836 & 29.0537 & 9.95547 \\
 0.0  & 1.43589 & 3.91027 & 1.55785 & 2.60235 & 31.9754 & 12.2871 \\
 0.1  & 1.33015 & 3.9499  & 1.41032 & 2.23319 & 34.7641 & 15.567 \\
 0.2  & 1.16733 & 3.98745 & 1.2     & 1.71993 & 37.2858 & 21.6786 \\
		\end{tabular}
	\end{ruledtabular}
\end{table*}
\endgroup
The black hole shadow's oblateness depends on the spin, the LSB parameter $\ell$, and the $ \theta_0$ . At extreme spin values, the oblateness varies significantly with $\ell$, becoming more elongated or less circular for larger $\ell$, especially at high spin. For negative $\ell$, the oblateness approaches 1, making the shadow nearly circular regardless of spin or $ \theta_0$  (see Figure~\ref{fig6B}). At higher inclination angles, the oblateness deviates more from 1, while at lower angles, it remains close to 1.

The shadow structure (Figure~\ref{fig2}) demonstrates that both $a$ and $\ell$ significantly impact the shadow area and oblateness. Observing only one shadow observable, either $A$ or $D$, can lead to ambiguities in parameter estimation. However, using both observables ($A, D$) allows for accurate determination of at least two parameters, as shown in Table \ref{table2} and Figure~\ref{contour2}. Confirming these observables enables precise estimation of the spin and $\ell$ values for a given mass and inclination.

An important finding is the relationship between the shadow area and the actual black hole size, represented by the event horizon area. As the LSB parameter $\ell$ increases, the event horizon area decreases while the shadow area increases. This counterintuitive result suggests that while the black hole's actual size shrinks, the perceived size of its shadow grows. This discrepancy provides insights into how the LSB parameter affects black hole spacetime, which profoundly impacts its geometric structure and gravitational projection into space.
\begingroup
\setlength{\tabcolsep}{8pt} % Default value: 6pt
\renewcommand{\arraystretch}{1} % Default value: 1
\begin{table*}
\caption{Summary of estimated constraints on the LSB parameter $\ell$ from Key observational tests.}\label{table5}
	\centering	
	\begin{tabular}
	 {p{6cm} |p{4cm}| p{5cm} }
		\hline
		Observational Tests  &  Estimated value of $\ell$ & References \\
		\hline\hline
Advance of perihelion & $10^{-8}-10^{-12}$ & Ref.~\citep{Casana:2017jkc} \\
Bending of light & $10^{-7}-10^{-15}$  &  Ref.~\citep{Casana:2017jkc}\\
Time delay of light & $10^{-9}-10^{-19}$  & Ref.~\citep{Casana:2017jkc} \\
GRO J1655-40    & $-0.1048_{-0.1316}^{+0.1678}$  & Ref.~\citep{Wang:2021gtd} \\
XTE J1550-564   & $-0.2053_{-0.3635}^{+6.7573}$  & Ref.~\citep{Wang:2021gtd} \\
GRS 1915+105    & $+1.3083_{-2.0134}^{+9.5717}$  & Ref.~\citep{Wang:2021gtd} \\
NuSTAR data of EXO 1846-031 & $0.46 \pm \text{B}$  &  Ref.~\citep{Gu:2022grg}\\
EHT DATA of M87* ($a=0.5M$) & $-0.5-2.87$  & Ref.~\citep{Wang:2021irh} \\
EHT DATA of M87* ($a=0.94M$) & $-0.5-0.132$  & Ref.~\citep{Wang:2021irh} \\
\end{tabular}
\end{table*}
\endgroup

\section{Constraints from the EHT Observation}\label{Sec-5}
A black hole shadow provides a direct diagnostic of strong-field gravity by revealing the black hole's influence on spacetime. This silhouette is created by the black hole's intense gravitational field bending and capturing light against the bright accretion disk \citep{Jaroszynski:1997bw,Falcke:1999pj}. The Event Horizon Telescope (EHT) has captured the first images of black hole shadows, including those of M87* and Sgr A*, enabling precise tests of gravitational theories \citep{EventHorizonTelescope:2019dse,EventHorizonTelescope:2022xnr}. These observations have tightly constrained the size and shape of black hole shadows, providing valuable data for testing General Relativity (GR) and alternative gravity theories in strong-field regimes. By comparing shadow observables with EHT data, we can explore black holes in Bumblebee gravity, enhancing our understanding of gravity under extreme conditions and potentially uncovering new physics beyond GR.
Previous attempts to constrain the Lorentz symmetry breaking (LSB) parameter include \citep{Wang:2021gtd,Gu:2022grg,Casana:2017jkc,Wang:2021irh}. Notably, Ref.~\citep{Wang:2021gtd} constrained black hole properties such as spin and the LSB parameter introduced by the bumblebee field, comparing Einstein-bumblebee theory predictions for quasi-periodic oscillation frequencies with observational data. Additionally, spectral data from X-ray emissions, such as the 2019 NuSTAR observation of the Galactic black hole EXO 1846-031, were used to test Einstein-bumblebee theory \citep{Gu:2022grg}. This analysis revealed a strong degeneracy between the LSB parameter $\ell$ and the black hole spin, indicating that variations in one could mimic changes in the other. This degeneracy underscores the need for additional data or methods to independently estimate the black hole spin and effectively test Lorentz symmetry breaking.

EHT observations can test black hole properties in Bumblebee gravity by analyzing shadow observables such as the shadow angular diameter, Schwarzschild radius deviation, and circularity deviation. High-resolution images of Sgr A* and M87* by EHT are crucial for these analyses. Deviations in the shadow angular diameter from GR predictions can reveal the influence of the vector field in Bumblebee gravity. The angular diameter $\theta_{sh}$ is defined as:
\begin{equation}\label{angularDiameterEq} \theta_{sh}=\frac{2}{d}\sqrt{\frac{A}{\pi}}, \end{equation}
where $A$ is the shadow area and $d$ is the distance from Earth.
Deviations from the Schwarzschild radius, representing the idealized size of a black hole shadow in Schwarzschild spacetime, can indicate deviations from general relativity. The Schwarzschild shadow deviation ($\delta$) measures the difference between the shadow angular diameter $\theta_{sh}$ of the rotating black hole in Bumblebee gravity and the diameter $\theta_{sh, Sch} = 6\sqrt{3}M$ of a Schwarzschild black hole. This deviation helps quantify how Lorentz-violating effects alter the black hole shadow and provides insights into modifications to spacetime structure caused by the LSB parameter $\ell$\citep{EventHorizonTelescope:2022xnr,EventHorizonTelescope:2022xqj}
\begin{equation}
\delta=\frac{\theta_{sh}}{6\sqrt{3}}-1.
\end{equation}
The Schwarzschild shadow deviation for a Kerr black hole with $a\leq M$ ranges from $-0.075$ to $0$ as the inclination varies from $0$ to $\pi/2$. The shadows of RBHBG can vary in size depending on the vector field and spacetime.

The circularity deviation, $\Delta C$, measures how much a black hole's shadow deviates from a perfect circle, offering insights into the symmetry of its gravitational field. The shadow boundary is described in polar coordinates as $(R(\varphi), \varphi)$, where $R(\varphi)$ is the radial distance to the boundary and $\varphi$ is the polar angle. The center of the shadow is at $(X_c, Y_c)$, with $X_c$ computed as $(X_r - X_l)/2$ and $Y_c = 0$ for symmetry along the Y-axis. The circularity deviation $\Delta C$ quantifies how much $R(\varphi)$ deviates from the average shadow radius $\bar{R}$, calculated as the root-mean-square deviation. The formula for $\Delta C$ is given by \citep{Johannsen:2010ru,Johannsen:2015qca, Kumar:2020yem}:
\begin{equation}\label{circular}
\Delta C=\frac{1}{\bar{R}}\sqrt{\frac{1}{\pi}\int_0^{2\pi}\left(R(\varphi)-\bar{R}\right)^2d\varphi},
\end{equation}
where $\bar{R}$ is the shadow average radius defined as \citep{Johannsen:2010ru}
\begin{equation}
\bar{R}=\frac{1}{2\pi}\int_{0}^{2\pi} R(\varphi) d\varphi.
\end{equation}
Here, $\varphi$ is the polar angle defined by $\varphi \equiv \tan^{-1}[Y/(X - X_c)]$, and $R(\varphi) = \sqrt{(X - X_c)^2 + (Y - Y_c)^2}$ is the radial distance from the shadow's center $(X_c, Y_c)$ to any boundary point $(X, Y)$. The observable $\Delta C$ measures the deviation of the shadow from a perfect circle. While a perfect circle has $\Delta C = 0$, deviations occur due to factors like the black hole's spin, the LSB parameter $\ell$ in Bumblebee gravity, or the inclination angle $\theta_0$. These deviations reveal how the black hole’s gravitational field and spacetime geometry influence the shadow's shape, providing insights into gravity in strong-field regimes and testing alternative theories involving Lorentz violation.
\begin{figure}
\begin{center}
	\includegraphics[scale=0.75]{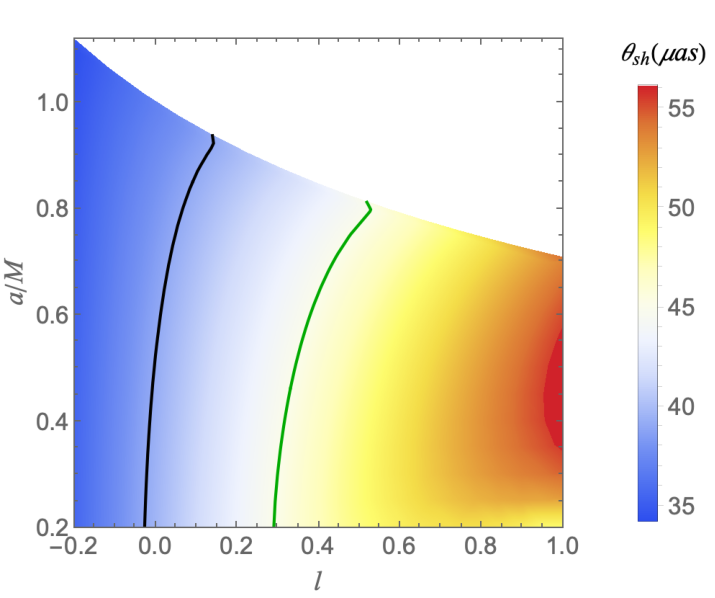}
	\caption{Figure shows the shadow angular diameter $\theta_{sh}$ of the RBHBG in $\mu$as as a function of the parameters $a/M$ and $l$ at an $ \theta_0$  of $17\degree$. The black line at $\theta_{sh} = 39\mu$as and the green line at $\theta_{sh} = 45\mu$as represent the bounds of the $1\sigma$ confidence region for the M87* shadow angular diameter, observed by the EHT as $\theta_{sh} = 42 \pm 3\mu$as. The parameter space within these lines corresponds to shadow sizes consistent with the EHT observations of M87*, while the white region indicates forbidden values for ($a/M, l/M$), where the shadow size does not match the observed constraints. }\label{M871}
\end{center}
\end{figure}

\begin{figure}
\begin{center}
	\includegraphics[scale=0.75]{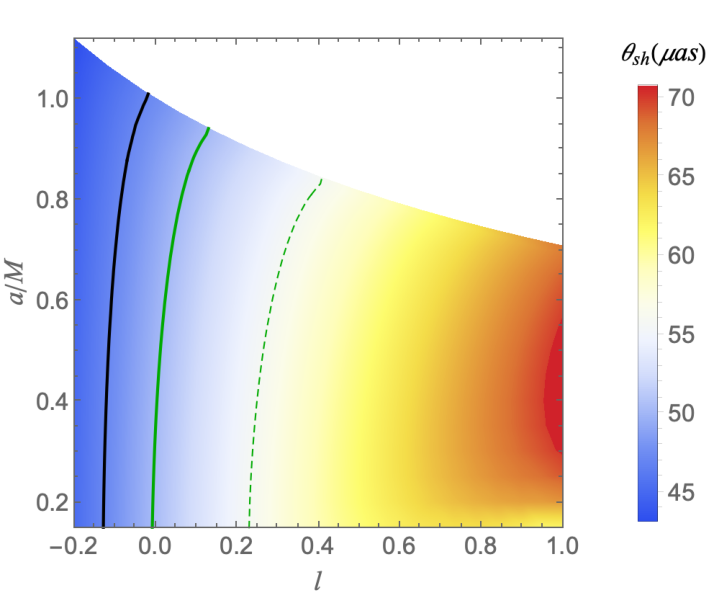}
	\caption{Figure illustrates the shadow angular diameter $\theta_{sh}$ of the RBHBG in $\mu$as as a function of the parameters $a/M$ and $l$ at an $ \theta_0$  of $50\degree$. The black line represents $\theta_{sh} = 50\mu$as, the green line corresponds to $\theta_{sh} = 46.9\mu$as and the dashed green corresponds to $\theta_{sh} =  55.7\mu$as. The region between these lines indicates parameter values where the black hole's shadow is consistent with the Sgr A* shadow size observed by the EHT. The white region denotes forbidden values for ($a/M, l/M$), where the shadow size is incompatible with the observed constraints. }\label{sgr1}
\end{center}	
\end{figure}

\paragraph{EHT constraints from M87*:} The EHT measurements of the shadow angular size for M87* have provided crucial insights into the nature of black holes \citep{EventHorizonTelescope:2019dse}.  Based on a priori known estimates for the mass and distance from stellar dynamics, these measurements were primarily compared against a large library of synthetic images generated from general-relativistic magnetohydrodynamics (GRMHD) simulations of accreting Kerr black holes. This extensive comparison enabled the EHT to derive a posterior distribution for the angular radius.  Through meticulous analysis, the EHT team was able to determine that the angular radius  of the shadow to be  approximately $\theta_{sh}=42 \pm 3 \mu as$.   The uncertainty is expressed with a 68\% confidence level \citep{EventHorizonTelescope:2019dse}. 
\begin{figure}
	\includegraphics[scale=0.75]{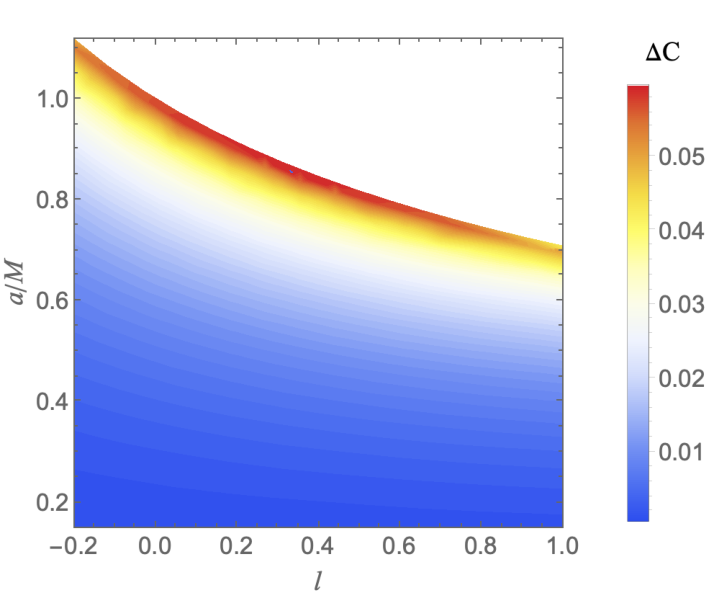}
\caption{Figure shows the deviation from circularity $\Delta C$ of the RBHBG, quantified as the root-mean-square distance from the average radius of the shadow. The EHT constraint of $\Delta C = 0.1$ is satisfied across all allowed parameter values where the black hole's shadow is consistent with the M87* shadow size observed by the EHT. The white region represents forbidden values for ($a/M, l$). }\label{deltaC}
\end{figure}

\begin{figure*}
    \begin{tabular}{p{9cm} p{9cm}}
     \includegraphics[scale=0.7]{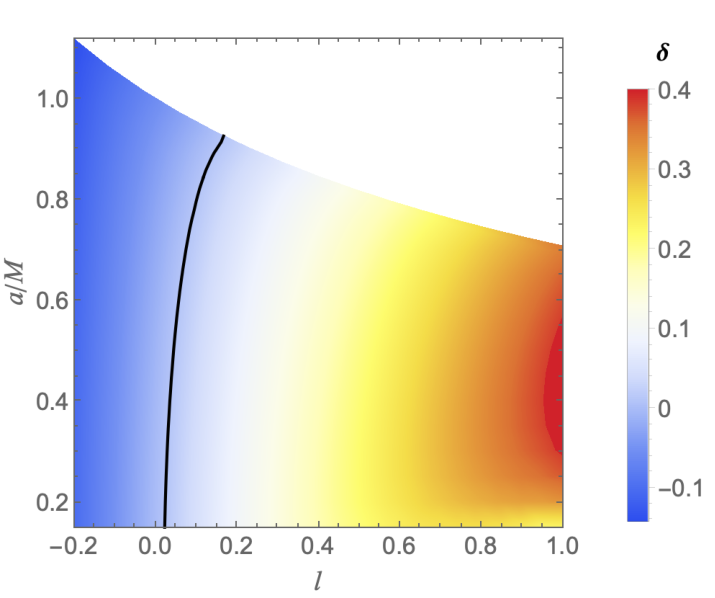}&
    \includegraphics[scale=0.7]{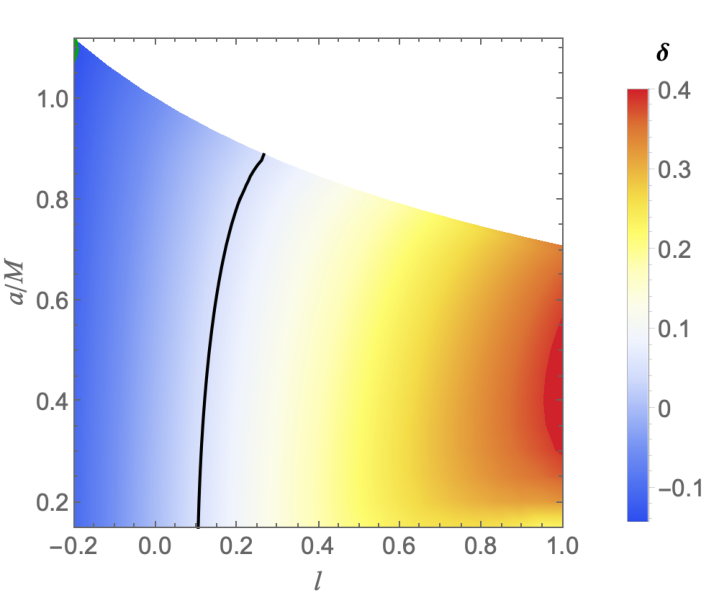}
    \end{tabular}
    \caption{Figure displays the deviation of the RBHBG shadow angular diameter from that of a Schwarzschild black hole as a function of the parameters ($a, l$). The constraints imposed by the EHT observations of the Sgr A* black hole shadow are shown by the VLTI bound at $\delta = 0.01$ (\textit{Left}) and the Keck bound at $\delta = 0.05$ (\textit{Right}). The white region represents forbidden values for ($a/M, l/M$).}\label{sgr2}
\end{figure*}
A rigorous comparison with non-Kerr black hole solutions would ideally require building similar libraries of synthetic images from GRMHD simulations specific to those non-Kerr models. However, creating these equivalent libraries is computationally unfeasible due to the vast parameter space and the complexity involved in such simulations. Moreover, the necessity of this approach is questionable in practice. Recent comparative analysis \citep{Mizuno:2018lxz,Vincent:2020dij} has demonstrated that the synthetic image libraries for Kerr and non-Kerr solutions would be very similar and essentially indistinguishable from the current observational quality. This similarity arises because the differences in the shadow angular sizes for Kerr and non-Kerr black holes are minimal compared to the observational uncertainties. Therefore, we adopt the working assumption that the $1-\sigma$ uncertainty in the shadow angular size for non-Kerr solutions is very similar to that for Kerr black holes. This allows us to employ the constraints derived for Kerr black holes for all the solutions considered. This approach simplifies the analysis while remaining consistent with the available observational data and theoretical predictions. Therefore, it is appropriate and timely to assess the viability of the RBHBG using the M87* black hole shadow observations. By examining the deviation from the circularity of the black hole shadow and the angular diameter, we establish constraints on the RBHBG parameters. This analysis confirms whether the RBHBG model is suitable for describing the M87* black hole, ensuring it matches the observed characteristics.

The parameter constraints for RBHBG, derived from the black hole shadow angular diameter measurements within the $1\sigma$ bound ($39\mu as \leq \theta_{sh} \leq 45\mu as$), reveal significant insights into the allowed values for the LSB parameter $l$. The angular diameter constraint places a bound on $l$ that varies with the spin parameter $a$; for example, at $a = 0.5 M$, $l$ is constrained to the range of $-0.0037$ to $0.1934$ (see Figure \ref{M871} and Table \ref{table3AA}). Notably, the EHT observations do not constrain the spin parameter $a$, allowing flexibility in its value. Within this constrained parameter space, the M87* black hole can be described by the RBHBG spacetime model, suggesting that these black holes are strong candidates for astrophysical black holes. Furthermore, as illustrated in Figure \ref{deltaC}, the circularity deviation bound $\Delta C \leq 0.10$ is satisfied across the entire parameter space for an inclination angle of $\theta_o = 17^\circ$, indicating that the shadows of RBHBG are nearly circular at small inclination angles, which aligns well with the observational constraints for M87*.

\paragraph{EHT constraints from Sgr A*:}
The EHT collaboration released the first image of Sgr A*, showing a compact emission region with variability on intrahour timescales \citep{EventHorizonTelescope:2022exc,EventHorizonTelescope:2022urf,EventHorizonTelescope:2022vjs,EventHorizonTelescope:2022wok,EventHorizonTelescope:2022xnr,EventHorizonTelescope:2022xqj}. Numerical simulations suggest this image matches the expected appearance of a Kerr black hole with a mass of about $4 \times 10^6 M$ and a distance of $8 \text{ kpc}$, aligning with precise astrometric measurements of S-star orbits \citep{Ghez:2003rt,Ghez:2008ms,Gillessen:2008qv}. EHT models indicate an inclination angle of around $30^\circ$ and a spin parameter $\chi > 0.5$ \citep{Fragione:2020khu}, ruling out high-inclination and non-spinning scenarios. Given its high curvature regime and accurate mass-to-distance ratio estimates, Sgr A* is an ideal target for testing astrophysical black hole models. We will use EHT data on the shadow angular diameter $\theta_{sh}$ and Schwarzschild shadow deviation $\delta$ to constrain the parameters of the RBHBG model.

The analysis of parameters for RBHBG, based on the black hole shadow angular diameter $\theta_{sh}$, shows that the range $41.7 \mu as \leq \theta_{sh} \leq 55.7 \mu as$, within the $1\sigma$ region for the Sgr A* shadow diameter, accommodates a broad parameter space for the RBHBG model (see Figure~\ref{sgr1}). However, more stringent constraints are provided by the EHT, which, using imaging algorithms \texttt{eht-imaging}, \texttt{SIMLI}, and \texttt{DIFMAP}, determines a narrower range of $46.9 \mu as \leq \theta_{sh} \leq 50 \mu as$. This tighter range imposes significant constraints on the LSB parameter $\ell$, whose extremal values depend on the spin parameter $a$. For instance, at $a = 0.5M$, $\ell$ is constrained between $-0.1162$ and $0.02267$, while at $a = 0.9M$, it ranges from $-0.08548$ to $0.1019$ ( cf. Table \ref{table3AA}). 

Furthermore, upper bounds on deviations from the Schwarzschild radius provided by the Keck and VLTI observations i.e., $\delta=-0.04^{+0.09}_{-0.01}$ and $\delta=-0.08^{+0.09}_{-0.09}$, respectively,   suggest that   $\ell<0.0436$ for $a = 0.5M$ and $\ell<0.1485$ for $a = 0.9M$  (cf.  Figure~\ref{sgr2} and Table \ref{table3AA}). Within these constrained parameter ranges, the RBHBG model's shadow is consistent with the Sgr A* shadow observed by the EHT.

\begin{table*}[htb!]
\caption{Maximum and minimum values of the LSB parameter for different spin values derived from shadow observables of M87* and Sgr A* using EHT observations.}\label{table3AA}
 \begin{centering}	
	\begin{tabular}{p{3cm}| p{1.2cm} p{1.2cm} p{1.2cm} p{1.2cm} p{1.2cm} p{1.2cm} p{1.2cm} p{1.2cm}}
\hline\hline
\text{EHT Observations} & {$a/M$ } & $0.4$ & $0.5$ & $0.6$ &$0.7$ & $0.8$  & $0.9$ &  $0.95$ \\
 \hline \hline

\multirow{2}{*}{$\theta_{\text{M87*}}$}  & $\ell_{min}$ & -0.01457 & -0.0037 & 0.01156 & 0.0334 & 0.0665 & 0.125 \\

                                 &$\ell_{max}$  & 0.1489 & 0.1672 & 0.1934 & 0.2321 & 0.2952 & 0.4567 \\
 \hline
 
\multirow{2}{*}{$\theta_{\text{Sgr A*}}$}  & $\ell_{min}$ &  -0.1215 & -0.1162 & -0.1089 & -0.09913 & -0.08548 & -0.06489 \\

                                 &$\ell_{max}$  & 0.00288 & 0.01123 & 0.02267 & 0.0385 & 0.06169 & 0.1019 \\               
\hline       
\multirow{2}{*}{$\delta_{\text{Keck}}$}  & $\ell_{min}$ &  - & - &  -  &  - &  - &  - &  -   \\ 

                                 &$\ell_{max}$  &   0.0345 & 0.0436 & 0.0563 & 0.07395 & 0.100064 & 0.14845 \\                       
 \hline
\multirow{2}{*}{$\delta_{\text{VLTI}}$}  & $\ell_{min}$ &  -0.08835 & -0.08225 & -0.07398 & -0.0627 & -0.0468 & -0.02218 \\

                                 &$\ell_{max}$  &  0.1198 & 0.1314 & 0.14754 & 0.17045 & 0.20586 & 0.2963 \\              

\hline  \hline       
                       
	\end{tabular}
\end{centering}
\end{table*} 

\section{Conclusion}\label{Sec-6}
We have investigated the properties of rotating black holes within the framework of Bumblebee gravity- namely RBHBG - wherein Lorentz symmetry is spontaneously broken by the vacuum expectation value of a vector field, mainly focusing on the impact of the LSB parameter $\ell$. By using high-resolution EHT observations images of Sgr A* and M87*, we have analyzed how deviations from GR manifest in the shadow characteristics of these black holes. Our findings indicate that the RBHBG model introduces notable deviations from the Kerr black hole scenario. Specifically, the parameter $\ell$ is found to increase the shadow radius and enhance deformation while reducing the event horizon area. This study enhances our understanding of the effects of black hole rotation and Lorentz symmetry breaking, offering insights into modified gravity theories and contributing to reconciling GR with quantum gravity.

We have demonstrated that in the RBHBG model, unlike the Kerr black hole, the maximum spin parameter $a$ can exceed the black hole mass $M$ when $\ell<0$. The event horizon radii decrease with increasing spin $a$ for all values of $\ell$. Our calculations of photon orbit radii show that as $\ell$ increases, the radius of retrograde photon orbits $r_{p}^{+}$ increases, while the radius of prograde photon orbits $r_{p}^{-}$ decreases. Comparing RBHBG to Kerr black holes at the same spin value, $r_{p}^{+}$ is smaller for $\ell<0$ and larger otherwise, while $r_{p}^{-}$ is larger for $\ell<0$ and smaller otherwise. For all $\ell$, the photon orbit radii in RBHBG remain within the ranges $M \leq r_{p}^{-} \leq 3M$ and $3M \leq r_{p}^{+} \leq 4M$.

We employ two established techniques for parameter estimation using shadow observables, allowing us to infer black hole parameters from observational data. Our analysis shows that the parameter $\ell$ consistently enlarges the shadow radius, indicating significant alterations to the black hole's spacetime. Additionally, $\ell$ influences shadow distortion mainly at higher inclination angles and spins, with less effect at lower angles. The shadow area exhibits distinct dependencies on $\ell$, spin, and inclination angle. 

Both $\ell$ and spin are crucial, with their effects modulated by the inclination angle $\theta_0$ , showing more pronounced differences at lower angles. The oblateness of the shadow varies notably with $\ell$ at extreme spin values: for larger $\ell$ and fixed inclination angles, it becomes significantly less than 1, making the shadow more elongated. For negative $\ell$, oblateness approaches 1 regardless of spin or inclination. The key finding is that as $\ell$ increases, the event horizon area decreases while the shadow area increases. These results highlight new avenues for studying black hole properties and Lorentz symmetry violations. Understanding the relationship between a black hole's actual size and its shadow could improve interpretations of observational data and refine black hole models. It underscores the importance of considering the LSB parameter in black hole metrics and its potential to transform our understanding of black hole dynamics.

We further modeled M87* and Sgr A* as RBHBG, using observational data from the EHT to test black hole properties by examining shadow observables such as shadow angular diameter, Schwarzschild radius deviation, and circularity deviation.  Within the $1\sigma$ bound of $39\mu as\leq\theta_{sh} \leq 45\mu as$, we observed that the shadow angular diameter for the M87* black hole provides a bound on $\ell$ that varies with the spin parameter $a$; for $a = 0.5 M$, for instance, $l$ is constrained to the range of $-0.0037$ to $0.1934$. The condition $\Delta C\leq 0.10$ is met by RBHBG for all values in the parameter space of M87*. This is because the shadows of the revolving black holes are almost perfectly circular when observed from small inclination angles.  

Additionally,  more stringent constraints are provided by the EHT, which, using imaging algorithms \texttt{eht-imaging}, \texttt{SIMLI}, and \texttt{DIFMAP}, and determines a narrower range of $46.9 \mu as \leq \theta_{sh} \leq 50 \mu as$ at an inclination angle of $\theta_o=50$\textdegree. This tighter range imposes significant constraints on the LSB parameter $l$, with its extremal values depending on the spin parameter $a$. For instance, at $a = 0.5M$, $\ell$ is constrained between $-0.1162$ and $0.02267$, while at $a = 0.9M$, it ranges from $-0.08548$ to $0.1019$.  Furthermore, upper bounds are obtained using the results on  deviations from the Schwarzschild radius provided by the Keck and VLTI observations which  suggest that   $\ell<0.0436$ for $a = 0.5M$ (Keck) and $\ell<0.1485$ for $a = 0.9M$ (VLTI). These results highlight the constraints that EHT observations place on the parameters of RBHBG, providing new insights into the properties of black holes and the nature of Lorentz symmetry violations in gravitational theories.

Our study highlights the importance of the LSB parameter in black hole metrics, demonstrating its potential to reshape our understanding of black hole dynamics. Incorporating this parameter reveals how deviations from standard gravitational theories can impact astrophysical observations, enhancing our insight into black hole behaviour. This approach improves our understanding of black hole physics and contributes to evaluating quantum gravity theories and the nature of spacetime at the Planck scale, offering a clearer view of the universe's fundamental structure.

\section{Acknowledgments} 
S.U.I would like to thank the University of KwaZulu-Natal and the NRF for the postdoctoral research fellowship. S.G.G. is supported by SERB-DST through project No. CRG/2021/005771. S.D.M acknowledges that this work is based upon research supported by the South African Research Chair Initiative of the Department of Science and Technology and the National Research Foundation.
\bibliography{Bumbelbee}

\begin{thebibliography}{85}%
\makeatletter
\providecommand \@ifxundefined [1]{%
 \@ifx{#1\undefined}
}%
\providecommand \@ifnum [1]{%
 \ifnum #1\expandafter \@firstoftwo
 \else \expandafter \@secondoftwo
 \fi
}%
\providecommand \@ifx [1]{%
 \ifx #1\expandafter \@firstoftwo
 \else \expandafter \@secondoftwo
 \fi
}%
\providecommand \natexlab [1]{#1}%
\providecommand \enquote  [1]{``#1''}%
\providecommand \bibnamefont  [1]{#1}%
\providecommand \bibfnamefont [1]{#1}%
\providecommand \citenamefont [1]{#1}%
\providecommand \href@noop [0]{\@secondoftwo}%
\providecommand \href [0]{\begingroup \@sanitize@url \@href}%
\providecommand \@href[1]{\@@startlink{#1}\@@href}%
\providecommand \@@href[1]{\endgroup#1\@@endlink}%
\providecommand \@sanitize@url [0]{\catcode `\\12\catcode `\$12\catcode `\&12\catcode `\#12\catcode `\^12\catcode `\_12\catcode `\%12\relax}%
\providecommand \@@startlink[1]{}%
\providecommand \@@endlink[0]{}%
\providecommand \url  [0]{\begingroup\@sanitize@url \@url }%
\providecommand \@url [1]{\endgroup\@href {#1}{\urlprefix }}%
\providecommand \urlprefix  [0]{URL }%
\providecommand \Eprint [0]{\href }%
\providecommand \doibase [0]{http://dx.doi.org/}%
\providecommand \selectlanguage [0]{\@gobble}%
\providecommand \bibinfo  [0]{\@secondoftwo}%
\providecommand \bibfield  [0]{\@secondoftwo}%
\providecommand \translation [1]{[#1]}%
\providecommand \BibitemOpen [0]{}%
\providecommand \bibitemStop [0]{}%
\providecommand \bibitemNoStop [0]{.\EOS\space}%
\providecommand \EOS [0]{\spacefactor3000\relax}%
\providecommand \BibitemShut  [1]{\csname bibitem#1\endcsname}%
\let\auto@bib@innerbib\@empty
%</preamble>
\bibitem [{\citenamefont {Griffiths}(2008)}]{Griffiths:2008zz}%
  \BibitemOpen
  \bibfield  {author} {\bibinfo {author} {\bibfnamefont {D.}~\bibnamefont {Griffiths}},\ }\href@noop {} {\emph {\bibinfo {title} {{Introduction to elementary particles}}}}\ (\bibinfo  {publisher} {Wiley-VCH},\ \bibinfo {year} {2008})\BibitemShut {NoStop}%
\bibitem [{\citenamefont {Rovelli}(2004)}]{Rovelli:2004tv}%
  \BibitemOpen
  \bibfield  {author} {\bibinfo {author} {\bibfnamefont {C.}~\bibnamefont {Rovelli}},\ }\href {\doibase 10.1017/CBO9780511755804} {\emph {\bibinfo {title} {{Quantum gravity}}}},\ Cambridge Monographs on Mathematical Physics\ (\bibinfo  {publisher} {Univ. Pr.},\ \bibinfo {address} {Cambridge, UK},\ \bibinfo {year} {2004})\BibitemShut {NoStop}%
\bibitem [{\citenamefont {Schutz}(1985)}]{Schutz:1985jx}%
  \BibitemOpen
  \bibfield  {author} {\bibinfo {author} {\bibfnamefont {B.~F.}\ \bibnamefont {Schutz}},\ }\href {\doibase 10.1017/CBO9780511984181} {\emph {\bibinfo {title} {{A FIRST COURSE IN GENERAL RELATIVITY}}}}\ (\bibinfo  {publisher} {Cambridge Univ. Pr.},\ \bibinfo {address} {Cambridge, UK},\ \bibinfo {year} {1985})\BibitemShut {NoStop}%
\bibitem [{\citenamefont {Liberati}(2013)}]{Liberati:2013xla}%
  \BibitemOpen
  \bibfield  {author} {\bibinfo {author} {\bibfnamefont {S.}~\bibnamefont {Liberati}},\ }\href {\doibase 10.1088/0264-9381/30/13/133001} {\bibfield  {journal} {\bibinfo  {journal} {Class. Quant. Grav.}\ }\textbf {\bibinfo {volume} {30}},\ \bibinfo {pages} {133001} (\bibinfo {year} {2013})},\ \Eprint {http://arxiv.org/abs/1304.5795} {arXiv:1304.5795 [gr-qc]} \BibitemShut {NoStop}%
\bibitem [{\citenamefont {Mattingly}(2005)}]{Mattingly:2005re}%
  \BibitemOpen
  \bibfield  {author} {\bibinfo {author} {\bibfnamefont {D.}~\bibnamefont {Mattingly}},\ }\href {\doibase 10.12942/lrr-2005-5} {\bibfield  {journal} {\bibinfo  {journal} {Living Rev. Rel.}\ }\textbf {\bibinfo {volume} {8}},\ \bibinfo {pages} {5} (\bibinfo {year} {2005})},\ \Eprint {http://arxiv.org/abs/gr-qc/0502097} {arXiv:gr-qc/0502097} \BibitemShut {NoStop}%
\bibitem [{\citenamefont {Kostelecky}(2004)}]{Kostelecky:2003fs}%
  \BibitemOpen
  \bibfield  {author} {\bibinfo {author} {\bibfnamefont {V.~A.}\ \bibnamefont {Kostelecky}},\ }\href {\doibase 10.1103/PhysRevD.69.105009} {\bibfield  {journal} {\bibinfo  {journal} {Phys. Rev. D}\ }\textbf {\bibinfo {volume} {69}},\ \bibinfo {pages} {105009} (\bibinfo {year} {2004})},\ \Eprint {http://arxiv.org/abs/hep-th/0312310} {arXiv:hep-th/0312310} \BibitemShut {NoStop}%
\bibitem [{\citenamefont {Bluhm}(2007)}]{Bluhm:2007xzd}%
  \BibitemOpen
  \bibfield  {author} {\bibinfo {author} {\bibfnamefont {R.}~\bibnamefont {Bluhm}},\ }\href {\doibase 10.22323/1.043.0009} {\bibfield  {journal} {\bibinfo  {journal} {PoS}\ }\textbf {\bibinfo {volume} {QG-PH}},\ \bibinfo {pages} {009} (\bibinfo {year} {2007})},\ \Eprint {http://arxiv.org/abs/0801.0141} {arXiv:0801.0141 [gr-qc]} \BibitemShut {NoStop}%
\bibitem [{\citenamefont {Kostelecky}\ and\ \citenamefont {Samuel}(1989{\natexlab{a}})}]{Kostelecky:1988zi}%
  \BibitemOpen
  \bibfield  {author} {\bibinfo {author} {\bibfnamefont {V.~A.}\ \bibnamefont {Kostelecky}}\ and\ \bibinfo {author} {\bibfnamefont {S.}~\bibnamefont {Samuel}},\ }\href {\doibase 10.1103/PhysRevD.39.683} {\bibfield  {journal} {\bibinfo  {journal} {Phys. Rev. D}\ }\textbf {\bibinfo {volume} {39}},\ \bibinfo {pages} {683} (\bibinfo {year} {1989}{\natexlab{a}})}\BibitemShut {NoStop}%
\bibitem [{\citenamefont {Kostelecky}\ and\ \citenamefont {Samuel}(1989{\natexlab{b}})}]{Kostelecky:1989jp}%
  \BibitemOpen
  \bibfield  {author} {\bibinfo {author} {\bibfnamefont {V.~A.}\ \bibnamefont {Kostelecky}}\ and\ \bibinfo {author} {\bibfnamefont {S.}~\bibnamefont {Samuel}},\ }\href {\doibase 10.1103/PhysRevLett.63.224} {\bibfield  {journal} {\bibinfo  {journal} {Phys. Rev. Lett.}\ }\textbf {\bibinfo {volume} {63}},\ \bibinfo {pages} {224} (\bibinfo {year} {1989}{\natexlab{b}})}\BibitemShut {NoStop}%
\bibitem [{\citenamefont {Carroll}\ \emph {et~al.}(2001)\citenamefont {Carroll}, \citenamefont {Harvey}, \citenamefont {Kostelecky}, \citenamefont {Lane},\ and\ \citenamefont {Okamoto}}]{Carroll:2001ws}%
  \BibitemOpen
  \bibfield  {author} {\bibinfo {author} {\bibfnamefont {S.~M.}\ \bibnamefont {Carroll}}, \bibinfo {author} {\bibfnamefont {J.~A.}\ \bibnamefont {Harvey}}, \bibinfo {author} {\bibfnamefont {V.~A.}\ \bibnamefont {Kostelecky}}, \bibinfo {author} {\bibfnamefont {C.~D.}\ \bibnamefont {Lane}}, \ and\ \bibinfo {author} {\bibfnamefont {T.}~\bibnamefont {Okamoto}},\ }\href {\doibase 10.1103/PhysRevLett.87.141601} {\bibfield  {journal} {\bibinfo  {journal} {Phys. Rev. Lett.}\ }\textbf {\bibinfo {volume} {87}},\ \bibinfo {pages} {141601} (\bibinfo {year} {2001})},\ \Eprint {http://arxiv.org/abs/hep-th/0105082} {arXiv:hep-th/0105082} \BibitemShut {NoStop}%
\bibitem [{\citenamefont {Gambini}\ and\ \citenamefont {Pullin}(1999)}]{Gambini:1998it}%
  \BibitemOpen
  \bibfield  {author} {\bibinfo {author} {\bibfnamefont {R.}~\bibnamefont {Gambini}}\ and\ \bibinfo {author} {\bibfnamefont {J.}~\bibnamefont {Pullin}},\ }\href {\doibase 10.1103/PhysRevD.59.124021} {\bibfield  {journal} {\bibinfo  {journal} {Phys. Rev. D}\ }\textbf {\bibinfo {volume} {59}},\ \bibinfo {pages} {124021} (\bibinfo {year} {1999})},\ \Eprint {http://arxiv.org/abs/gr-qc/9809038} {arXiv:gr-qc/9809038} \BibitemShut {NoStop}%
\bibitem [{\citenamefont {Kostelecky}\ and\ \citenamefont {Samuel}(1989{\natexlab{c}})}]{Kostelecky:1989jw}%
  \BibitemOpen
  \bibfield  {author} {\bibinfo {author} {\bibfnamefont {V.~A.}\ \bibnamefont {Kostelecky}}\ and\ \bibinfo {author} {\bibfnamefont {S.}~\bibnamefont {Samuel}},\ }\href {\doibase 10.1103/PhysRevD.40.1886} {\bibfield  {journal} {\bibinfo  {journal} {Phys. Rev. D}\ }\textbf {\bibinfo {volume} {40}},\ \bibinfo {pages} {1886} (\bibinfo {year} {1989}{\natexlab{c}})}\BibitemShut {NoStop}%
\bibitem [{\citenamefont {Bluhm}\ and\ \citenamefont {Kostelecky}(2005)}]{Bluhm:2004ep}%
  \BibitemOpen
  \bibfield  {author} {\bibinfo {author} {\bibfnamefont {R.}~\bibnamefont {Bluhm}}\ and\ \bibinfo {author} {\bibfnamefont {V.~A.}\ \bibnamefont {Kostelecky}},\ }\href {\doibase 10.1103/PhysRevD.71.065008} {\bibfield  {journal} {\bibinfo  {journal} {Phys. Rev. D}\ }\textbf {\bibinfo {volume} {71}},\ \bibinfo {pages} {065008} (\bibinfo {year} {2005})},\ \Eprint {http://arxiv.org/abs/hep-th/0412320} {arXiv:hep-th/0412320} \BibitemShut {NoStop}%
\bibitem [{\citenamefont {Bertolami}\ and\ \citenamefont {Paramos}(2005)}]{Bertolami:2005bh}%
  \BibitemOpen
  \bibfield  {author} {\bibinfo {author} {\bibfnamefont {O.}~\bibnamefont {Bertolami}}\ and\ \bibinfo {author} {\bibfnamefont {J.}~\bibnamefont {Paramos}},\ }\href {\doibase 10.1103/PhysRevD.72.044001} {\bibfield  {journal} {\bibinfo  {journal} {Phys. Rev. D}\ }\textbf {\bibinfo {volume} {72}},\ \bibinfo {pages} {044001} (\bibinfo {year} {2005})},\ \Eprint {http://arxiv.org/abs/hep-th/0504215} {arXiv:hep-th/0504215} \BibitemShut {NoStop}%
\bibitem [{\citenamefont {Bailey}\ and\ \citenamefont {Kostelecky}(2006)}]{Bailey:2006fd}%
  \BibitemOpen
  \bibfield  {author} {\bibinfo {author} {\bibfnamefont {Q.~G.}\ \bibnamefont {Bailey}}\ and\ \bibinfo {author} {\bibfnamefont {V.~A.}\ \bibnamefont {Kostelecky}},\ }\href {\doibase 10.1103/PhysRevD.74.045001} {\bibfield  {journal} {\bibinfo  {journal} {Phys. Rev. D}\ }\textbf {\bibinfo {volume} {74}},\ \bibinfo {pages} {045001} (\bibinfo {year} {2006})},\ \Eprint {http://arxiv.org/abs/gr-qc/0603030} {arXiv:gr-qc/0603030} \BibitemShut {NoStop}%
\bibitem [{\citenamefont {Bluhm}\ \emph {et~al.}(2008)\citenamefont {Bluhm}, \citenamefont {Gagne}, \citenamefont {Potting},\ and\ \citenamefont {Vrublevskis}}]{Bluhm:2008yt}%
  \BibitemOpen
  \bibfield  {author} {\bibinfo {author} {\bibfnamefont {R.}~\bibnamefont {Bluhm}}, \bibinfo {author} {\bibfnamefont {N.~L.}\ \bibnamefont {Gagne}}, \bibinfo {author} {\bibfnamefont {R.}~\bibnamefont {Potting}}, \ and\ \bibinfo {author} {\bibfnamefont {A.}~\bibnamefont {Vrublevskis}},\ }\href {\doibase 10.1103/PhysRevD.79.029902} {\bibfield  {journal} {\bibinfo  {journal} {Phys. Rev. D}\ }\textbf {\bibinfo {volume} {77}},\ \bibinfo {pages} {125007} (\bibinfo {year} {2008})},\ \bibinfo {note} {[Erratum: Phys.Rev.D 79, 029902 (2009)]},\ \Eprint {http://arxiv.org/abs/0802.4071} {arXiv:0802.4071 [hep-th]} \BibitemShut {NoStop}%
\bibitem [{\citenamefont {Seifert}(2010)}]{Seifert:2009gi}%
  \BibitemOpen
  \bibfield  {author} {\bibinfo {author} {\bibfnamefont {M.~D.}\ \bibnamefont {Seifert}},\ }\href {\doibase 10.1103/PhysRevD.81.065010} {\bibfield  {journal} {\bibinfo  {journal} {Phys. Rev. D}\ }\textbf {\bibinfo {volume} {81}},\ \bibinfo {pages} {065010} (\bibinfo {year} {2010})},\ \Eprint {http://arxiv.org/abs/0909.3118} {arXiv:0909.3118 [hep-ph]} \BibitemShut {NoStop}%
\bibitem [{\citenamefont {Maluf}\ \emph {et~al.}(2014)\citenamefont {Maluf}, \citenamefont {Almeida}, \citenamefont {Casana},\ and\ \citenamefont {Ferreira}}]{Maluf:2014dpa}%
  \BibitemOpen
  \bibfield  {author} {\bibinfo {author} {\bibfnamefont {R.~V.}\ \bibnamefont {Maluf}}, \bibinfo {author} {\bibfnamefont {C.~A.~S.}\ \bibnamefont {Almeida}}, \bibinfo {author} {\bibfnamefont {R.}~\bibnamefont {Casana}}, \ and\ \bibinfo {author} {\bibfnamefont {M.~M.}\ \bibnamefont {Ferreira}, \bibfnamefont {Jr.}},\ }\href {\doibase 10.1103/PhysRevD.90.025007} {\bibfield  {journal} {\bibinfo  {journal} {Phys. Rev. D}\ }\textbf {\bibinfo {volume} {90}},\ \bibinfo {pages} {025007} (\bibinfo {year} {2014})},\ \Eprint {http://arxiv.org/abs/1402.3554} {arXiv:1402.3554 [hep-th]} \BibitemShut {NoStop}%
\bibitem [{\citenamefont {P\'aramos}\ and\ \citenamefont {Guiomar}(2014)}]{Paramos:2014mda}%
  \BibitemOpen
  \bibfield  {author} {\bibinfo {author} {\bibfnamefont {J.}~\bibnamefont {P\'aramos}}\ and\ \bibinfo {author} {\bibfnamefont {G.}~\bibnamefont {Guiomar}},\ }\href {\doibase 10.1103/PhysRevD.90.082002} {\bibfield  {journal} {\bibinfo  {journal} {Phys. Rev. D}\ }\textbf {\bibinfo {volume} {90}},\ \bibinfo {pages} {082002} (\bibinfo {year} {2014})},\ \Eprint {http://arxiv.org/abs/1409.2022} {arXiv:1409.2022 [astro-ph.SR]} \BibitemShut {NoStop}%
\bibitem [{\citenamefont {Assun\c{c}\~ao}\ \emph {et~al.}(2019)\citenamefont {Assun\c{c}\~ao}, \citenamefont {Mariz}, \citenamefont {Nascimento},\ and\ \citenamefont {Petrov}}]{Assuncao:2019azw}%
  \BibitemOpen
  \bibfield  {author} {\bibinfo {author} {\bibfnamefont {J.~F.}\ \bibnamefont {Assun\c{c}\~ao}}, \bibinfo {author} {\bibfnamefont {T.}~\bibnamefont {Mariz}}, \bibinfo {author} {\bibfnamefont {J.~R.}\ \bibnamefont {Nascimento}}, \ and\ \bibinfo {author} {\bibfnamefont {A.~Y.}\ \bibnamefont {Petrov}},\ }\href {\doibase 10.1103/PhysRevD.100.085009} {\bibfield  {journal} {\bibinfo  {journal} {Phys. Rev. D}\ }\textbf {\bibinfo {volume} {100}},\ \bibinfo {pages} {085009} (\bibinfo {year} {2019})},\ \Eprint {http://arxiv.org/abs/1902.10592} {arXiv:1902.10592 [hep-th]} \BibitemShut {NoStop}%
\bibitem [{\citenamefont {Escobar}\ and\ \citenamefont {Mart\'\i{}n-Ruiz}(2017)}]{Escobar:2017fdi}%
  \BibitemOpen
  \bibfield  {author} {\bibinfo {author} {\bibfnamefont {C.~A.}\ \bibnamefont {Escobar}}\ and\ \bibinfo {author} {\bibfnamefont {A.}~\bibnamefont {Mart\'\i{}n-Ruiz}},\ }\href {\doibase 10.1103/PhysRevD.95.095006} {\bibfield  {journal} {\bibinfo  {journal} {Phys. Rev. D}\ }\textbf {\bibinfo {volume} {95}},\ \bibinfo {pages} {095006} (\bibinfo {year} {2017})},\ \Eprint {http://arxiv.org/abs/1703.01171} {arXiv:1703.01171 [hep-th]} \BibitemShut {NoStop}%
\bibitem [{\citenamefont {Casana}\ \emph {et~al.}(2018)\citenamefont {Casana}, \citenamefont {Cavalcante}, \citenamefont {Poulis},\ and\ \citenamefont {Santos}}]{Casana:2017jkc}%
  \BibitemOpen
  \bibfield  {author} {\bibinfo {author} {\bibfnamefont {R.}~\bibnamefont {Casana}}, \bibinfo {author} {\bibfnamefont {A.}~\bibnamefont {Cavalcante}}, \bibinfo {author} {\bibfnamefont {F.~P.}\ \bibnamefont {Poulis}}, \ and\ \bibinfo {author} {\bibfnamefont {E.~B.}\ \bibnamefont {Santos}},\ }\href {\doibase 10.1103/PhysRevD.97.104001} {\bibfield  {journal} {\bibinfo  {journal} {Phys. Rev. D}\ }\textbf {\bibinfo {volume} {97}},\ \bibinfo {pages} {104001} (\bibinfo {year} {2018})},\ \Eprint {http://arxiv.org/abs/1711.02273} {arXiv:1711.02273 [gr-qc]} \BibitemShut {NoStop}%
\bibitem [{\citenamefont {Ovg\"un}\ \emph {et~al.}(2018)\citenamefont {Ovg\"un}, \citenamefont {Jusufi},\ and\ \citenamefont {Sakalli}}]{Ovgun:2018ran}%
  \BibitemOpen
  \bibfield  {author} {\bibinfo {author} {\bibfnamefont {A.}~\bibnamefont {Ovg\"un}}, \bibinfo {author} {\bibfnamefont {K.}~\bibnamefont {Jusufi}}, \ and\ \bibinfo {author} {\bibfnamefont {I.}~\bibnamefont {Sakalli}},\ }\href {\doibase 10.1016/j.aop.2018.10.012} {\bibfield  {journal} {\bibinfo  {journal} {Annals Phys.}\ }\textbf {\bibinfo {volume} {399}},\ \bibinfo {pages} {193} (\bibinfo {year} {2018})},\ \Eprint {http://arxiv.org/abs/1805.09431} {arXiv:1805.09431 [gr-qc]} \BibitemShut {NoStop}%
\bibitem [{\citenamefont {Oliveira}\ \emph {et~al.}(2021)\citenamefont {Oliveira}, \citenamefont {Dantas},\ and\ \citenamefont {Almeida}}]{Oliveira:2021abg}%
  \BibitemOpen
  \bibfield  {author} {\bibinfo {author} {\bibfnamefont {R.}~\bibnamefont {Oliveira}}, \bibinfo {author} {\bibfnamefont {D.~M.}\ \bibnamefont {Dantas}}, \ and\ \bibinfo {author} {\bibfnamefont {C.~A.~S.}\ \bibnamefont {Almeida}},\ }\href {\doibase 10.1209/0295-5075/ac130c} {\bibfield  {journal} {\bibinfo  {journal} {EPL}\ }\textbf {\bibinfo {volume} {135}},\ \bibinfo {pages} {10003} (\bibinfo {year} {2021})},\ \Eprint {http://arxiv.org/abs/2105.07956} {arXiv:2105.07956 [gr-qc]} \BibitemShut {NoStop}%
\bibitem [{\citenamefont {Kanzi}\ and\ \citenamefont {Sakall\i{}}(2019)}]{Kanzi:2019gtu}%
  \BibitemOpen
  \bibfield  {author} {\bibinfo {author} {\bibfnamefont {S.}~\bibnamefont {Kanzi}}\ and\ \bibinfo {author} {\bibfnamefont {I.}~\bibnamefont {Sakall\i{}}},\ }\href {\doibase 10.1016/j.nuclphysb.2019.114703} {\bibfield  {journal} {\bibinfo  {journal} {Nucl. Phys. B}\ }\textbf {\bibinfo {volume} {946}},\ \bibinfo {pages} {114703} (\bibinfo {year} {2019})},\ \Eprint {http://arxiv.org/abs/1905.00477} {arXiv:1905.00477 [hep-th]} \BibitemShut {NoStop}%
\bibitem [{\citenamefont {G\"ull\"u}\ and\ \citenamefont {\"Ovg\"un}(2022)}]{Gullu:2020qzu}%
  \BibitemOpen
  \bibfield  {author} {\bibinfo {author} {\bibfnamefont {I.}~\bibnamefont {G\"ull\"u}}\ and\ \bibinfo {author} {\bibfnamefont {A.}~\bibnamefont {\"Ovg\"un}},\ }\href {\doibase 10.1016/j.aop.2021.168721} {\bibfield  {journal} {\bibinfo  {journal} {Annals Phys.}\ }\textbf {\bibinfo {volume} {436}},\ \bibinfo {pages} {168721} (\bibinfo {year} {2022})},\ \Eprint {http://arxiv.org/abs/2012.02611} {arXiv:2012.02611 [gr-qc]} \BibitemShut {NoStop}%
\bibitem [{\citenamefont {Maluf}\ and\ \citenamefont {Neves}(2021)}]{Maluf:2020kgf}%
  \BibitemOpen
  \bibfield  {author} {\bibinfo {author} {\bibfnamefont {R.~V.}\ \bibnamefont {Maluf}}\ and\ \bibinfo {author} {\bibfnamefont {J.~C.~S.}\ \bibnamefont {Neves}},\ }\href {\doibase 10.1103/PhysRevD.103.044002} {\bibfield  {journal} {\bibinfo  {journal} {Phys. Rev. D}\ }\textbf {\bibinfo {volume} {103}},\ \bibinfo {pages} {044002} (\bibinfo {year} {2021})},\ \Eprint {http://arxiv.org/abs/2011.12841} {arXiv:2011.12841 [gr-qc]} \BibitemShut {NoStop}%
\bibitem [{\citenamefont {Ding}\ \emph {et~al.}(2022)\citenamefont {Ding}, \citenamefont {Chen},\ and\ \citenamefont {Fu}}]{Ding:2021iwv}%
  \BibitemOpen
  \bibfield  {author} {\bibinfo {author} {\bibfnamefont {C.}~\bibnamefont {Ding}}, \bibinfo {author} {\bibfnamefont {X.}~\bibnamefont {Chen}}, \ and\ \bibinfo {author} {\bibfnamefont {X.}~\bibnamefont {Fu}},\ }\href {\doibase 10.1016/j.nuclphysb.2022.115688} {\bibfield  {journal} {\bibinfo  {journal} {Nucl. Phys. B}\ }\textbf {\bibinfo {volume} {975}},\ \bibinfo {pages} {115688} (\bibinfo {year} {2022})},\ \Eprint {http://arxiv.org/abs/2102.13335} {arXiv:2102.13335 [gr-qc]} \BibitemShut {NoStop}%
\bibitem [{\citenamefont {\"Ovg\"un}\ \emph {et~al.}(2019)\citenamefont {\"Ovg\"un}, \citenamefont {Jusufi},\ and\ \citenamefont {Sakall\i{}}}]{Ovgun:2018xys}%
  \BibitemOpen
  \bibfield  {author} {\bibinfo {author} {\bibfnamefont {A.}~\bibnamefont {\"Ovg\"un}}, \bibinfo {author} {\bibfnamefont {K.}~\bibnamefont {Jusufi}}, \ and\ \bibinfo {author} {\bibfnamefont {I.}~\bibnamefont {Sakall\i{}}},\ }\href {\doibase 10.1103/PhysRevD.99.024042} {\bibfield  {journal} {\bibinfo  {journal} {Phys. Rev. D}\ }\textbf {\bibinfo {volume} {99}},\ \bibinfo {pages} {024042} (\bibinfo {year} {2019})},\ \Eprint {http://arxiv.org/abs/1804.09911} {arXiv:1804.09911 [gr-qc]} \BibitemShut {NoStop}%
\bibitem [{\citenamefont {Capelo}\ and\ \citenamefont {P\'aramos}(2015)}]{Capelo:2015ipa}%
  \BibitemOpen
  \bibfield  {author} {\bibinfo {author} {\bibfnamefont {D.}~\bibnamefont {Capelo}}\ and\ \bibinfo {author} {\bibfnamefont {J.}~\bibnamefont {P\'aramos}},\ }\href {\doibase 10.1103/PhysRevD.91.104007} {\bibfield  {journal} {\bibinfo  {journal} {Phys. Rev. D}\ }\textbf {\bibinfo {volume} {91}},\ \bibinfo {pages} {104007} (\bibinfo {year} {2015})},\ \Eprint {http://arxiv.org/abs/1501.07685} {arXiv:1501.07685 [gr-qc]} \BibitemShut {NoStop}%
\bibitem [{\citenamefont {Ding}\ \emph {et~al.}(2020)\citenamefont {Ding}, \citenamefont {Liu}, \citenamefont {Casana},\ and\ \citenamefont {Cavalcante}}]{Ding:2019mal}%
  \BibitemOpen
  \bibfield  {author} {\bibinfo {author} {\bibfnamefont {C.}~\bibnamefont {Ding}}, \bibinfo {author} {\bibfnamefont {C.}~\bibnamefont {Liu}}, \bibinfo {author} {\bibfnamefont {R.}~\bibnamefont {Casana}}, \ and\ \bibinfo {author} {\bibfnamefont {A.}~\bibnamefont {Cavalcante}},\ }\href {\doibase 10.1140/epjc/s10052-020-7743-y} {\bibfield  {journal} {\bibinfo  {journal} {Eur. Phys. J. C}\ }\textbf {\bibinfo {volume} {80}},\ \bibinfo {pages} {178} (\bibinfo {year} {2020})},\ \Eprint {http://arxiv.org/abs/1910.02674} {arXiv:1910.02674 [gr-qc]} \BibitemShut {NoStop}%
\bibitem [{\citenamefont {Akiyama}\ \emph {et~al.}(2019{\natexlab{a}})\citenamefont {Akiyama} \emph {et~al.}}]{EventHorizonTelescope:2019dse}%
  \BibitemOpen
  \bibfield  {author} {\bibinfo {author} {\bibfnamefont {K.}~\bibnamefont {Akiyama}} \emph {et~al.} (\bibinfo {collaboration} {Event Horizon Telescope}),\ }\href {\doibase 10.3847/2041-8213/ab0ec7} {\bibfield  {journal} {\bibinfo  {journal} {Astrophys. J. Lett.}\ }\textbf {\bibinfo {volume} {875}},\ \bibinfo {pages} {L1} (\bibinfo {year} {2019}{\natexlab{a}})},\ \Eprint {http://arxiv.org/abs/1906.11238} {arXiv:1906.11238 [astro-ph.GA]} \BibitemShut {NoStop}%
\bibitem [{\citenamefont {Wang}\ and\ \citenamefont {Wei}(2022)}]{Wang:2021irh}%
  \BibitemOpen
  \bibfield  {author} {\bibinfo {author} {\bibfnamefont {H.-M.}\ \bibnamefont {Wang}}\ and\ \bibinfo {author} {\bibfnamefont {S.-W.}\ \bibnamefont {Wei}},\ }\href {\doibase 10.1140/epjp/s13360-022-02785-6} {\bibfield  {journal} {\bibinfo  {journal} {Eur. Phys. J. Plus}\ }\textbf {\bibinfo {volume} {137}},\ \bibinfo {pages} {571} (\bibinfo {year} {2022})},\ \Eprint {http://arxiv.org/abs/2106.14602} {arXiv:2106.14602 [gr-qc]} \BibitemShut {NoStop}%
\bibitem [{\citenamefont {Liu}\ \emph {et~al.}(2019)\citenamefont {Liu}, \citenamefont {Ding},\ and\ \citenamefont {Jing}}]{Liu:2019mls}%
  \BibitemOpen
  \bibfield  {author} {\bibinfo {author} {\bibfnamefont {C.}~\bibnamefont {Liu}}, \bibinfo {author} {\bibfnamefont {C.}~\bibnamefont {Ding}}, \ and\ \bibinfo {author} {\bibfnamefont {J.}~\bibnamefont {Jing}},\ }\href@noop {} {\bibfield  {journal} {\bibinfo  {journal} {.s}\ } (\bibinfo {year} {2019})},\ \Eprint {http://arxiv.org/abs/1910.13259} {arXiv:1910.13259 [gr-qc]} \BibitemShut {NoStop}%
\bibitem [{\citenamefont {Jiang}\ \emph {et~al.}(2021)\citenamefont {Jiang}, \citenamefont {Lin},\ and\ \citenamefont {Zhai}}]{Jiang:2021whw}%
  \BibitemOpen
  \bibfield  {author} {\bibinfo {author} {\bibfnamefont {R.}~\bibnamefont {Jiang}}, \bibinfo {author} {\bibfnamefont {R.-H.}\ \bibnamefont {Lin}}, \ and\ \bibinfo {author} {\bibfnamefont {X.-H.}\ \bibnamefont {Zhai}},\ }\href {\doibase 10.1103/PhysRevD.104.124004} {\bibfield  {journal} {\bibinfo  {journal} {Phys. Rev. D}\ }\textbf {\bibinfo {volume} {104}},\ \bibinfo {pages} {124004} (\bibinfo {year} {2021})},\ \Eprint {http://arxiv.org/abs/2108.04702} {arXiv:2108.04702 [gr-qc]} \BibitemShut {NoStop}%
\bibitem [{\citenamefont {Li}\ and\ \citenamefont {\"Ovg\"un}(2020)}]{Li:2020dln}%
  \BibitemOpen
  \bibfield  {author} {\bibinfo {author} {\bibfnamefont {Z.}~\bibnamefont {Li}}\ and\ \bibinfo {author} {\bibfnamefont {A.}~\bibnamefont {\"Ovg\"un}},\ }\href {\doibase 10.1103/PhysRevD.101.024040} {\bibfield  {journal} {\bibinfo  {journal} {Phys. Rev. D}\ }\textbf {\bibinfo {volume} {101}},\ \bibinfo {pages} {024040} (\bibinfo {year} {2020})},\ \Eprint {http://arxiv.org/abs/2001.02074} {arXiv:2001.02074 [gr-qc]} \BibitemShut {NoStop}%
\bibitem [{\citenamefont {Wang}\ \emph {et~al.}(2022)\citenamefont {Wang}, \citenamefont {Chen},\ and\ \citenamefont {Jing}}]{Wang:2021gtd}%
  \BibitemOpen
  \bibfield  {author} {\bibinfo {author} {\bibfnamefont {Z.}~\bibnamefont {Wang}}, \bibinfo {author} {\bibfnamefont {S.}~\bibnamefont {Chen}}, \ and\ \bibinfo {author} {\bibfnamefont {J.}~\bibnamefont {Jing}},\ }\href {\doibase 10.1140/epjc/s10052-022-10475-x} {\bibfield  {journal} {\bibinfo  {journal} {Eur. Phys. J. C}\ }\textbf {\bibinfo {volume} {82}},\ \bibinfo {pages} {528} (\bibinfo {year} {2022})},\ \Eprint {http://arxiv.org/abs/2112.02895} {arXiv:2112.02895 [gr-qc]} \BibitemShut {NoStop}%
\bibitem [{\citenamefont {Gu}\ \emph {et~al.}(2022)\citenamefont {Gu}, \citenamefont {Riaz}, \citenamefont {Abdikamalov}, \citenamefont {Ayzenberg},\ and\ \citenamefont {Bambi}}]{Gu:2022grg}%
  \BibitemOpen
  \bibfield  {author} {\bibinfo {author} {\bibfnamefont {J.}~\bibnamefont {Gu}}, \bibinfo {author} {\bibfnamefont {S.}~\bibnamefont {Riaz}}, \bibinfo {author} {\bibfnamefont {A.~B.}\ \bibnamefont {Abdikamalov}}, \bibinfo {author} {\bibfnamefont {D.}~\bibnamefont {Ayzenberg}}, \ and\ \bibinfo {author} {\bibfnamefont {C.}~\bibnamefont {Bambi}},\ }\href {\doibase 10.1140/epjc/s10052-022-10686-2} {\bibfield  {journal} {\bibinfo  {journal} {Eur. Phys. J. C}\ }\textbf {\bibinfo {volume} {82}},\ \bibinfo {pages} {708} (\bibinfo {year} {2022})},\ \Eprint {http://arxiv.org/abs/2206.14733} {arXiv:2206.14733 [gr-qc]} \BibitemShut {NoStop}%
\bibitem [{\citenamefont {Azreg-A\"\i{}nou}(2014{\natexlab{a}})}]{Azreg-Ainou:2014pra}%
  \BibitemOpen
  \bibfield  {author} {\bibinfo {author} {\bibfnamefont {M.}~\bibnamefont {Azreg-A\"\i{}nou}},\ }\href {\doibase 10.1103/PhysRevD.90.064041} {\bibfield  {journal} {\bibinfo  {journal} {Phys. Rev. D}\ }\textbf {\bibinfo {volume} {90}},\ \bibinfo {pages} {064041} (\bibinfo {year} {2014}{\natexlab{a}})},\ \Eprint {http://arxiv.org/abs/1405.2569} {arXiv:1405.2569 [gr-qc]} \BibitemShut {NoStop}%
\bibitem [{\citenamefont {Azreg-A\"\i{}nou}(2014{\natexlab{b}})}]{Azreg-Ainou:2014aqa}%
  \BibitemOpen
  \bibfield  {author} {\bibinfo {author} {\bibfnamefont {M.}~\bibnamefont {Azreg-A\"\i{}nou}},\ }\href {\doibase 10.1140/epjc/s10052-014-2865-8} {\bibfield  {journal} {\bibinfo  {journal} {Eur. Phys. J. C}\ }\textbf {\bibinfo {volume} {74}},\ \bibinfo {pages} {2865} (\bibinfo {year} {2014}{\natexlab{b}})},\ \Eprint {http://arxiv.org/abs/1401.4292} {arXiv:1401.4292 [gr-qc]} \BibitemShut {NoStop}%
\bibitem [{\citenamefont {Newman}\ and\ \citenamefont {Janis}(1965)}]{Newman:1965tw}%
  \BibitemOpen
  \bibfield  {author} {\bibinfo {author} {\bibfnamefont {E.~T.}\ \bibnamefont {Newman}}\ and\ \bibinfo {author} {\bibfnamefont {A.~I.}\ \bibnamefont {Janis}},\ }\href {\doibase 10.1063/1.1704350} {\bibfield  {journal} {\bibinfo  {journal} {J. Math. Phys.}\ }\textbf {\bibinfo {volume} {6}},\ \bibinfo {pages} {915} (\bibinfo {year} {1965})}\BibitemShut {NoStop}%
\bibitem [{\citenamefont {Synge}(1966)}]{Synge:1966}%
  \BibitemOpen
  \bibfield  {author} {\bibinfo {author} {\bibfnamefont {J.~L.}\ \bibnamefont {Synge}},\ }\href {\doibase 10.1093/mnras/131.3.463} {\bibfield  {journal} {\bibinfo  {journal} {Monthly Notices of the Royal Astronomical Society}\ }\textbf {\bibinfo {volume} {131}},\ \bibinfo {pages} {463} (\bibinfo {year} {1966})},\ \Eprint {http://arxiv.org/abs/https://academic.oup.com/mnras/article-pdf/131/3/463/8076763/mnras131-0463.pdf} {https://academic.oup.com/mnras/article-pdf/131/3/463/8076763/mnras131-0463.pdf} \BibitemShut {NoStop}%
\bibitem [{\citenamefont {Bardeen}(1973)}]{bardeen1973}%
  \BibitemOpen
  \bibfield  {author} {\bibinfo {author} {\bibfnamefont {J.}~\bibnamefont {Bardeen}},\ }\href@noop {} {\enquote {\bibinfo {title} {Black holes, edited by c. dewitt and bs dewitt},}\ } (\bibinfo {year} {1973})\BibitemShut {NoStop}%
\bibitem [{\citenamefont {Luminet}(1979)}]{Luminet:1979}%
  \BibitemOpen
  \bibfield  {author} {\bibinfo {author} {\bibfnamefont {J.~P.}\ \bibnamefont {Luminet}},\ }\href@noop {} {\bibfield  {journal} {\bibinfo  {journal} {aap}\ }\textbf {\bibinfo {volume} {75}},\ \bibinfo {pages} {228} (\bibinfo {year} {1979})}\BibitemShut {NoStop}%
\bibitem [{\citenamefont {{Cunningham}}\ and\ \citenamefont {{Bardeen}}(1973)}]{Cunningham:1973}%
  \BibitemOpen
  \bibfield  {author} {\bibinfo {author} {\bibfnamefont {C.~T.}\ \bibnamefont {{Cunningham}}}\ and\ \bibinfo {author} {\bibfnamefont {J.~M.}\ \bibnamefont {{Bardeen}}},\ }\href {\doibase 10.1086/152223} {\bibfield  {journal} {\bibinfo  {journal} {\apj}\ }\textbf {\bibinfo {volume} {183}},\ \bibinfo {pages} {237} (\bibinfo {year} {1973})}\BibitemShut {NoStop}%
\bibitem [{\citenamefont {{de Vries}}(2000)}]{Vries:2000}%
  \BibitemOpen
  \bibfield  {author} {\bibinfo {author} {\bibfnamefont {A.}~\bibnamefont {{de Vries}}},\ }\href {\doibase 10.1088/0264-9381/17/1/309} {\bibfield  {journal} {\bibinfo  {journal} {Classical and Quantum Gravity}\ }\textbf {\bibinfo {volume} {17}},\ \bibinfo {pages} {123} (\bibinfo {year} {2000})}\BibitemShut {NoStop}%
\bibitem [{\citenamefont {Shen}\ \emph {et~al.}(2005)\citenamefont {Shen}, \citenamefont {Lo}, \citenamefont {Liang}, \citenamefont {Ho},\ and\ \citenamefont {Zhao}}]{Shen:2005cw}%
  \BibitemOpen
  \bibfield  {author} {\bibinfo {author} {\bibfnamefont {Z.-Q.}\ \bibnamefont {Shen}}, \bibinfo {author} {\bibfnamefont {K.~Y.}\ \bibnamefont {Lo}}, \bibinfo {author} {\bibfnamefont {M.~C.}\ \bibnamefont {Liang}}, \bibinfo {author} {\bibfnamefont {P.~T.~P.}\ \bibnamefont {Ho}}, \ and\ \bibinfo {author} {\bibfnamefont {J.~H.}\ \bibnamefont {Zhao}},\ }\href {\doibase 10.1038/nature04205} {\bibfield  {journal} {\bibinfo  {journal} {Nature}\ }\textbf {\bibinfo {volume} {438}},\ \bibinfo {pages} {62} (\bibinfo {year} {2005})},\ \Eprint {http://arxiv.org/abs/astro-ph/0512515} {arXiv:astro-ph/0512515} \BibitemShut {NoStop}%
\bibitem [{\citenamefont {Amarilla}\ \emph {et~al.}(2010)\citenamefont {Amarilla}, \citenamefont {Eiroa},\ and\ \citenamefont {Giribet}}]{Amarilla:2010zq}%
  \BibitemOpen
  \bibfield  {author} {\bibinfo {author} {\bibfnamefont {L.}~\bibnamefont {Amarilla}}, \bibinfo {author} {\bibfnamefont {E.~F.}\ \bibnamefont {Eiroa}}, \ and\ \bibinfo {author} {\bibfnamefont {G.}~\bibnamefont {Giribet}},\ }\href {\doibase 10.1103/PhysRevD.81.124045} {\bibfield  {journal} {\bibinfo  {journal} {Phys. Rev. D}\ }\textbf {\bibinfo {volume} {81}},\ \bibinfo {pages} {124045} (\bibinfo {year} {2010})},\ \Eprint {http://arxiv.org/abs/1005.0607} {arXiv:1005.0607 [gr-qc]} \BibitemShut {NoStop}%
\bibitem [{\citenamefont {Yumoto}\ \emph {et~al.}(2012)\citenamefont {Yumoto}, \citenamefont {Nitta}, \citenamefont {Chiba},\ and\ \citenamefont {Sugiyama}}]{Yumoto:2012kz}%
  \BibitemOpen
  \bibfield  {author} {\bibinfo {author} {\bibfnamefont {A.}~\bibnamefont {Yumoto}}, \bibinfo {author} {\bibfnamefont {D.}~\bibnamefont {Nitta}}, \bibinfo {author} {\bibfnamefont {T.}~\bibnamefont {Chiba}}, \ and\ \bibinfo {author} {\bibfnamefont {N.}~\bibnamefont {Sugiyama}},\ }\href {\doibase 10.1103/PhysRevD.86.103001} {\bibfield  {journal} {\bibinfo  {journal} {Phys. Rev. D}\ }\textbf {\bibinfo {volume} {86}},\ \bibinfo {pages} {103001} (\bibinfo {year} {2012})},\ \Eprint {http://arxiv.org/abs/1208.0635} {arXiv:1208.0635 [gr-qc]} \BibitemShut {NoStop}%
\bibitem [{\citenamefont {Amarilla}\ and\ \citenamefont {Eiroa}(2013)}]{Amarilla:2013sj}%
  \BibitemOpen
  \bibfield  {author} {\bibinfo {author} {\bibfnamefont {L.}~\bibnamefont {Amarilla}}\ and\ \bibinfo {author} {\bibfnamefont {E.~F.}\ \bibnamefont {Eiroa}},\ }\href {\doibase 10.1103/PhysRevD.87.044057} {\bibfield  {journal} {\bibinfo  {journal} {Phys. Rev. D}\ }\textbf {\bibinfo {volume} {87}},\ \bibinfo {pages} {044057} (\bibinfo {year} {2013})},\ \Eprint {http://arxiv.org/abs/1301.0532} {arXiv:1301.0532 [gr-qc]} \BibitemShut {NoStop}%
\bibitem [{\citenamefont {Atamurotov}\ \emph {et~al.}(2013)\citenamefont {Atamurotov}, \citenamefont {Abdujabbarov},\ and\ \citenamefont {Ahmedov}}]{Atamurotov:2013sca}%
  \BibitemOpen
  \bibfield  {author} {\bibinfo {author} {\bibfnamefont {F.}~\bibnamefont {Atamurotov}}, \bibinfo {author} {\bibfnamefont {A.}~\bibnamefont {Abdujabbarov}}, \ and\ \bibinfo {author} {\bibfnamefont {B.}~\bibnamefont {Ahmedov}},\ }\href {\doibase 10.1103/PhysRevD.88.064004} {\bibfield  {journal} {\bibinfo  {journal} {Phys. Rev. D}\ }\textbf {\bibinfo {volume} {88}},\ \bibinfo {pages} {064004} (\bibinfo {year} {2013})}\BibitemShut {NoStop}%
\bibitem [{\citenamefont {Abdujabbarov}\ \emph {et~al.}(2016)\citenamefont {Abdujabbarov}, \citenamefont {Amir}, \citenamefont {Ahmedov},\ and\ \citenamefont {Ghosh}}]{Abdujabbarov:2016hnw}%
  \BibitemOpen
  \bibfield  {author} {\bibinfo {author} {\bibfnamefont {A.}~\bibnamefont {Abdujabbarov}}, \bibinfo {author} {\bibfnamefont {M.}~\bibnamefont {Amir}}, \bibinfo {author} {\bibfnamefont {B.}~\bibnamefont {Ahmedov}}, \ and\ \bibinfo {author} {\bibfnamefont {S.~G.}\ \bibnamefont {Ghosh}},\ }\href {\doibase 10.1103/PhysRevD.93.104004} {\bibfield  {journal} {\bibinfo  {journal} {Phys. Rev.}\ }\textbf {\bibinfo {volume} {D93}},\ \bibinfo {pages} {104004} (\bibinfo {year} {2016})},\ \Eprint {http://arxiv.org/abs/1604.03809} {arXiv:1604.03809 [gr-qc]} \BibitemShut {NoStop}%
%%CITATION = ARXIV:1604.03809;%%
\bibitem [{\citenamefont {Abdujabbarov}\ \emph {et~al.}(2015)\citenamefont {Abdujabbarov}, \citenamefont {Rezzolla},\ and\ \citenamefont {Ahmedov}}]{Abdujabbarov:2015xqa}%
  \BibitemOpen
  \bibfield  {author} {\bibinfo {author} {\bibfnamefont {A.~A.}\ \bibnamefont {Abdujabbarov}}, \bibinfo {author} {\bibfnamefont {L.}~\bibnamefont {Rezzolla}}, \ and\ \bibinfo {author} {\bibfnamefont {B.~J.}\ \bibnamefont {Ahmedov}},\ }\href {\doibase 10.1093/mnras/stv2079} {\bibfield  {journal} {\bibinfo  {journal} {Mon. Not. Roy. Astron. Soc.}\ }\textbf {\bibinfo {volume} {454}},\ \bibinfo {pages} {2423} (\bibinfo {year} {2015})},\ \Eprint {http://arxiv.org/abs/1503.09054} {arXiv:1503.09054 [gr-qc]} \BibitemShut {NoStop}%
\bibitem [{\citenamefont {Cunha}\ and\ \citenamefont {Herdeiro}(2018)}]{Cunha:2018acu}%
  \BibitemOpen
  \bibfield  {author} {\bibinfo {author} {\bibfnamefont {P.~V.~P.}\ \bibnamefont {Cunha}}\ and\ \bibinfo {author} {\bibfnamefont {C.~A.~R.}\ \bibnamefont {Herdeiro}},\ }\href {\doibase 10.1007/s10714-018-2361-9} {\bibfield  {journal} {\bibinfo  {journal} {Gen. Rel. Grav.}\ }\textbf {\bibinfo {volume} {50}},\ \bibinfo {pages} {42} (\bibinfo {year} {2018})},\ \Eprint {http://arxiv.org/abs/1801.00860} {arXiv:1801.00860 [gr-qc]} \BibitemShut {NoStop}%
\bibitem [{\citenamefont {Mizuno}\ \emph {et~al.}(2018)\citenamefont {Mizuno}, \citenamefont {Younsi}, \citenamefont {Fromm}, \citenamefont {Porth}, \citenamefont {De~Laurentis}, \citenamefont {Olivares}, \citenamefont {Falcke}, \citenamefont {Kramer},\ and\ \citenamefont {Rezzolla}}]{Mizuno:2018lxz}%
  \BibitemOpen
  \bibfield  {author} {\bibinfo {author} {\bibfnamefont {Y.}~\bibnamefont {Mizuno}}, \bibinfo {author} {\bibfnamefont {Z.}~\bibnamefont {Younsi}}, \bibinfo {author} {\bibfnamefont {C.~M.}\ \bibnamefont {Fromm}}, \bibinfo {author} {\bibfnamefont {O.}~\bibnamefont {Porth}}, \bibinfo {author} {\bibfnamefont {M.}~\bibnamefont {De~Laurentis}}, \bibinfo {author} {\bibfnamefont {H.}~\bibnamefont {Olivares}}, \bibinfo {author} {\bibfnamefont {H.}~\bibnamefont {Falcke}}, \bibinfo {author} {\bibfnamefont {M.}~\bibnamefont {Kramer}}, \ and\ \bibinfo {author} {\bibfnamefont {L.}~\bibnamefont {Rezzolla}},\ }\href {\doibase 10.1038/s41550-018-0449-5} {\bibfield  {journal} {\bibinfo  {journal} {Nature Astron.}\ }\textbf {\bibinfo {volume} {2}},\ \bibinfo {pages} {585} (\bibinfo {year} {2018})},\ \Eprint {http://arxiv.org/abs/1804.05812} {arXiv:1804.05812 [astro-ph.GA]} \BibitemShut {NoStop}%
\bibitem [{\citenamefont {Mishra}\ \emph {et~al.}(2019)\citenamefont {Mishra}, \citenamefont {Chakraborty},\ and\ \citenamefont {Sarkar}}]{Mishra:2019trb}%
  \BibitemOpen
  \bibfield  {author} {\bibinfo {author} {\bibfnamefont {A.~K.}\ \bibnamefont {Mishra}}, \bibinfo {author} {\bibfnamefont {S.}~\bibnamefont {Chakraborty}}, \ and\ \bibinfo {author} {\bibfnamefont {S.}~\bibnamefont {Sarkar}},\ }\href {\doibase 10.1103/PhysRevD.99.104080} {\bibfield  {journal} {\bibinfo  {journal} {Phys. Rev. D}\ }\textbf {\bibinfo {volume} {99}},\ \bibinfo {pages} {104080} (\bibinfo {year} {2019})},\ \Eprint {http://arxiv.org/abs/1903.06376} {arXiv:1903.06376 [gr-qc]} \BibitemShut {NoStop}%
\bibitem [{\citenamefont {Shaikh}(2019)}]{Shaikh:2019fpu}%
  \BibitemOpen
  \bibfield  {author} {\bibinfo {author} {\bibfnamefont {R.}~\bibnamefont {Shaikh}},\ }\href {\doibase 10.1103/PhysRevD.100.024028} {\bibfield  {journal} {\bibinfo  {journal} {Phys. Rev. D}\ }\textbf {\bibinfo {volume} {100}},\ \bibinfo {pages} {024028} (\bibinfo {year} {2019})},\ \Eprint {http://arxiv.org/abs/1904.08322} {arXiv:1904.08322 [gr-qc]} \BibitemShut {NoStop}%
\bibitem [{\citenamefont {Kumar}\ \emph {et~al.}(2020)\citenamefont {Kumar}, \citenamefont {Kumar},\ and\ \citenamefont {Ghosh}}]{Kumar:2020yem}%
  \BibitemOpen
  \bibfield  {author} {\bibinfo {author} {\bibfnamefont {R.}~\bibnamefont {Kumar}}, \bibinfo {author} {\bibfnamefont {A.}~\bibnamefont {Kumar}}, \ and\ \bibinfo {author} {\bibfnamefont {S.~G.}\ \bibnamefont {Ghosh}},\ }\href {\doibase 10.3847/1538-4357/ab8c4a} {\bibfield  {journal} {\bibinfo  {journal} {Astrophys. J.}\ }\textbf {\bibinfo {volume} {896}},\ \bibinfo {pages} {89} (\bibinfo {year} {2020})},\ \Eprint {http://arxiv.org/abs/2006.09869} {arXiv:2006.09869 [gr-qc]} \BibitemShut {NoStop}%
\bibitem [{\citenamefont {Kumar}\ and\ \citenamefont {Ghosh}(2020)}]{Kumar:2018ple}%
  \BibitemOpen
  \bibfield  {author} {\bibinfo {author} {\bibfnamefont {R.}~\bibnamefont {Kumar}}\ and\ \bibinfo {author} {\bibfnamefont {S.~G.}\ \bibnamefont {Ghosh}},\ }\href {\doibase 10.3847/1538-4357/ab77b0} {\bibfield  {journal} {\bibinfo  {journal} {Astrophys. J.}\ }\textbf {\bibinfo {volume} {892}},\ \bibinfo {pages} {78} (\bibinfo {year} {2020})},\ \Eprint {http://arxiv.org/abs/1811.01260} {arXiv:1811.01260 [gr-qc]} \BibitemShut {NoStop}%
\bibitem [{\citenamefont {Kramer}\ \emph {et~al.}(2004)\citenamefont {Kramer}, \citenamefont {Backer}, \citenamefont {Cordes}, \citenamefont {Lazio}, \citenamefont {Stappers},\ and\ \citenamefont {Johnston}}]{Kramer:2004hd}%
  \BibitemOpen
  \bibfield  {author} {\bibinfo {author} {\bibfnamefont {M.}~\bibnamefont {Kramer}}, \bibinfo {author} {\bibfnamefont {D.~C.}\ \bibnamefont {Backer}}, \bibinfo {author} {\bibfnamefont {J.~M.}\ \bibnamefont {Cordes}}, \bibinfo {author} {\bibfnamefont {T.~J.~W.}\ \bibnamefont {Lazio}}, \bibinfo {author} {\bibfnamefont {B.~W.}\ \bibnamefont {Stappers}}, \ and\ \bibinfo {author} {\bibfnamefont {S.}~\bibnamefont {Johnston}},\ }\href {\doibase 10.1016/j.newar.2004.09.020} {\bibfield  {journal} {\bibinfo  {journal} {New Astron. Rev.}\ }\textbf {\bibinfo {volume} {48}},\ \bibinfo {pages} {993} (\bibinfo {year} {2004})},\ \Eprint {http://arxiv.org/abs/astro-ph/0409379} {arXiv:astro-ph/0409379} \BibitemShut {NoStop}%
\bibitem [{\citenamefont {Carter}(1968)}]{Carter:1968rr}%
  \BibitemOpen
  \bibfield  {author} {\bibinfo {author} {\bibfnamefont {B.}~\bibnamefont {Carter}},\ }\href {\doibase 10.1103/PhysRev.174.1559} {\bibfield  {journal} {\bibinfo  {journal} {Phys. Rev.}\ }\textbf {\bibinfo {volume} {174}},\ \bibinfo {pages} {1559} (\bibinfo {year} {1968})}\BibitemShut {NoStop}%
%%CITATION = PHRVA,174,1559;%%
\bibitem [{\citenamefont {Chandrasekhar}(1985)}]{Chandrasekhar:1985kt}%
  \BibitemOpen
  \bibfield  {author} {\bibinfo {author} {\bibfnamefont {S.}~\bibnamefont {Chandrasekhar}},\ }\href@noop {} {\emph {\bibinfo {title} {{The mathematical theory of black holes}}}}\ (\bibinfo  {publisher} {Oxford Univ. Press},\ \bibinfo {address} {Oxford},\ \bibinfo {year} {1985})\BibitemShut {NoStop}%
%%CITATION = INSPIRE-224457;%%
\bibitem [{\citenamefont {Hioki}\ and\ \citenamefont {Maeda}(2009)}]{Hioki:2009na}%
  \BibitemOpen
  \bibfield  {author} {\bibinfo {author} {\bibfnamefont {K.}~\bibnamefont {Hioki}}\ and\ \bibinfo {author} {\bibfnamefont {K.-i.}\ \bibnamefont {Maeda}},\ }\href {\doibase 10.1103/PhysRevD.80.024042} {\bibfield  {journal} {\bibinfo  {journal} {Phys. Rev. D}\ }\textbf {\bibinfo {volume} {80}},\ \bibinfo {pages} {024042} (\bibinfo {year} {2009})},\ \Eprint {http://arxiv.org/abs/0904.3575} {arXiv:0904.3575 [astro-ph.HE]} \BibitemShut {NoStop}%
\bibitem [{\citenamefont {{Israel}}(1967)}]{Werner:1967ab}%
  \BibitemOpen
  \bibfield  {author} {\bibinfo {author} {\bibfnamefont {W.}~\bibnamefont {{Israel}}},\ }\href {\doibase 10.1103/PhysRev.164.1776} {\bibfield  {journal} {\bibinfo  {journal} {Physical Review}\ }\textbf {\bibinfo {volume} {164}},\ \bibinfo {pages} {1776} (\bibinfo {year} {1967})}\BibitemShut {NoStop}%
\bibitem [{\citenamefont {{Israel}}(1968)}]{Werner:1968ab}%
  \BibitemOpen
  \bibfield  {author} {\bibinfo {author} {\bibfnamefont {W.}~\bibnamefont {{Israel}}},\ }\href {\doibase 10.1007/BF01645859} {\bibfield  {journal} {\bibinfo  {journal} {Communications in Mathematical Physics}\ }\textbf {\bibinfo {volume} {8}},\ \bibinfo {pages} {245} (\bibinfo {year} {1968})}\BibitemShut {NoStop}%
\bibitem [{\citenamefont {Carter}(1971)}]{Carter:1971zc}%
  \BibitemOpen
  \bibfield  {author} {\bibinfo {author} {\bibfnamefont {B.}~\bibnamefont {Carter}},\ }\href {\doibase 10.1103/PhysRevLett.26.331} {\bibfield  {journal} {\bibinfo  {journal} {Phys. Rev. Lett.}\ }\textbf {\bibinfo {volume} {26}},\ \bibinfo {pages} {331} (\bibinfo {year} {1971})}\BibitemShut {NoStop}%
\bibitem [{\citenamefont {Misner}\ \emph {et~al.}(1973)\citenamefont {Misner}, \citenamefont {Thorne},\ and\ \citenamefont {Wheeler}}]{Misner:1973prb}%
  \BibitemOpen
  \bibfield  {author} {\bibinfo {author} {\bibfnamefont {C.~W.}\ \bibnamefont {Misner}}, \bibinfo {author} {\bibfnamefont {K.~S.}\ \bibnamefont {Thorne}}, \ and\ \bibinfo {author} {\bibfnamefont {J.~A.}\ \bibnamefont {Wheeler}},\ }\href@noop {} {\emph {\bibinfo {title} {{Gravitation}}}}\ (\bibinfo  {publisher} {W. H. Freeman},\ \bibinfo {address} {San Francisco},\ \bibinfo {year} {1973})\BibitemShut {NoStop}%
\bibitem [{\citenamefont {Akiyama}\ \emph {et~al.}(2019{\natexlab{b}})\citenamefont {Akiyama} \emph {et~al.}}]{EventHorizonTelescope:2019ggy}%
  \BibitemOpen
  \bibfield  {author} {\bibinfo {author} {\bibfnamefont {K.}~\bibnamefont {Akiyama}} \emph {et~al.} (\bibinfo {collaboration} {Event Horizon Telescope}),\ }\href {\doibase 10.3847/2041-8213/ab1141} {\bibfield  {journal} {\bibinfo  {journal} {Astrophys. J. Lett.}\ }\textbf {\bibinfo {volume} {875}},\ \bibinfo {pages} {L6} (\bibinfo {year} {2019}{\natexlab{b}})},\ \Eprint {http://arxiv.org/abs/1906.11243} {arXiv:1906.11243 [astro-ph.GA]} \BibitemShut {NoStop}%
\bibitem [{\citenamefont {Chen}\ \emph {et~al.}(2019)\citenamefont {Chen} \emph {et~al.}}]{Chen:2019tdb}%
  \BibitemOpen
  \bibfield  {author} {\bibinfo {author} {\bibfnamefont {Z.}~\bibnamefont {Chen}} \emph {et~al.},\ }\href {\doibase 10.3847/2041-8213/ab3c68} {\bibfield  {journal} {\bibinfo  {journal} {Astrophys. J. Lett.}\ }\textbf {\bibinfo {volume} {882}},\ \bibinfo {pages} {L28} (\bibinfo {year} {2019})},\ \Eprint {http://arxiv.org/abs/1908.08066} {arXiv:1908.08066 [astro-ph.GA]} \BibitemShut {NoStop}%
\bibitem [{\citenamefont {Tsupko}(2017)}]{Tsupko:2017rdo}%
  \BibitemOpen
  \bibfield  {author} {\bibinfo {author} {\bibfnamefont {O.~{\relax Yu}.}\ \bibnamefont {Tsupko}},\ }\href {\doibase 10.1103/PhysRevD.95.104058} {\bibfield  {journal} {\bibinfo  {journal} {Phys. Rev.}\ }\textbf {\bibinfo {volume} {D95}},\ \bibinfo {pages} {104058} (\bibinfo {year} {2017})},\ \Eprint {http://arxiv.org/abs/1702.04005} {arXiv:1702.04005 [gr-qc]} \BibitemShut {NoStop}%
%%CITATION = ARXIV:1702.04005;%%
\bibitem [{\citenamefont {Jaroszynski}\ and\ \citenamefont {Kurpiewski}(1997)}]{Jaroszynski:1997bw}%
  \BibitemOpen
  \bibfield  {author} {\bibinfo {author} {\bibfnamefont {M.}~\bibnamefont {Jaroszynski}}\ and\ \bibinfo {author} {\bibfnamefont {A.}~\bibnamefont {Kurpiewski}},\ }\href@noop {} {\bibfield  {journal} {\bibinfo  {journal} {Astron. Astrophys.}\ }\textbf {\bibinfo {volume} {326}},\ \bibinfo {pages} {419} (\bibinfo {year} {1997})},\ \Eprint {http://arxiv.org/abs/astro-ph/9705044} {arXiv:astro-ph/9705044} \BibitemShut {NoStop}%
\bibitem [{\citenamefont {Falcke}\ \emph {et~al.}(2000)\citenamefont {Falcke}, \citenamefont {Melia},\ and\ \citenamefont {Agol}}]{Falcke:1999pj}%
  \BibitemOpen
  \bibfield  {author} {\bibinfo {author} {\bibfnamefont {H.}~\bibnamefont {Falcke}}, \bibinfo {author} {\bibfnamefont {F.}~\bibnamefont {Melia}}, \ and\ \bibinfo {author} {\bibfnamefont {E.}~\bibnamefont {Agol}},\ }\href {\doibase 10.1086/312423} {\bibfield  {journal} {\bibinfo  {journal} {Astrophys. J. Lett.}\ }\textbf {\bibinfo {volume} {528}},\ \bibinfo {pages} {L13} (\bibinfo {year} {2000})},\ \Eprint {http://arxiv.org/abs/astro-ph/9912263} {arXiv:astro-ph/9912263} \BibitemShut {NoStop}%
\bibitem [{\citenamefont {Akiyama}\ \emph {et~al.}(2022{\natexlab{a}})\citenamefont {Akiyama} \emph {et~al.}}]{EventHorizonTelescope:2022xnr}%
  \BibitemOpen
  \bibfield  {author} {\bibinfo {author} {\bibfnamefont {K.}~\bibnamefont {Akiyama}} \emph {et~al.} (\bibinfo {collaboration} {Event Horizon Telescope}),\ }\href {\doibase 10.3847/2041-8213/ac6674} {\bibfield  {journal} {\bibinfo  {journal} {Astrophys. J. Lett.}\ }\textbf {\bibinfo {volume} {930}},\ \bibinfo {pages} {L12} (\bibinfo {year} {2022}{\natexlab{a}})}\BibitemShut {NoStop}%
\bibitem [{\citenamefont {Akiyama}\ \emph {et~al.}(2022{\natexlab{b}})\citenamefont {Akiyama} \emph {et~al.}}]{EventHorizonTelescope:2022xqj}%
  \BibitemOpen
  \bibfield  {author} {\bibinfo {author} {\bibfnamefont {K.}~\bibnamefont {Akiyama}} \emph {et~al.} (\bibinfo {collaboration} {Event Horizon Telescope}),\ }\href {\doibase 10.3847/2041-8213/ac6756} {\bibfield  {journal} {\bibinfo  {journal} {Astrophys. J. Lett.}\ }\textbf {\bibinfo {volume} {930}},\ \bibinfo {pages} {L17} (\bibinfo {year} {2022}{\natexlab{b}})}\BibitemShut {NoStop}%
\bibitem [{\citenamefont {Johannsen}\ and\ \citenamefont {Psaltis}(2010)}]{Johannsen:2010ru}%
  \BibitemOpen
  \bibfield  {author} {\bibinfo {author} {\bibfnamefont {T.}~\bibnamefont {Johannsen}}\ and\ \bibinfo {author} {\bibfnamefont {D.}~\bibnamefont {Psaltis}},\ }\href {\doibase 10.1088/0004-637X/718/1/446} {\bibfield  {journal} {\bibinfo  {journal} {Astrophys. J.}\ }\textbf {\bibinfo {volume} {718}},\ \bibinfo {pages} {446} (\bibinfo {year} {2010})},\ \Eprint {http://arxiv.org/abs/1005.1931} {arXiv:1005.1931 [astro-ph.HE]} \BibitemShut {NoStop}%
%%CITATION = ARXIV:1005.1931;%%
\bibitem [{\citenamefont {Johannsen}(2013)}]{Johannsen:2015qca}%
  \BibitemOpen
  \bibfield  {author} {\bibinfo {author} {\bibfnamefont {T.}~\bibnamefont {Johannsen}},\ }\href {\doibase 10.1088/0004-637X/777/2/170} {\bibfield  {journal} {\bibinfo  {journal} {Astrophys. J.}\ }\textbf {\bibinfo {volume} {777}},\ \bibinfo {pages} {170} (\bibinfo {year} {2013})},\ \Eprint {http://arxiv.org/abs/1501.02814} {arXiv:1501.02814 [astro-ph.HE]} \BibitemShut {NoStop}%
%%CITATION = ARXIV:1501.02814;%%
\bibitem [{\citenamefont {Vincent}\ \emph {et~al.}(2021)\citenamefont {Vincent}, \citenamefont {Wielgus}, \citenamefont {Abramowicz}, \citenamefont {Gourgoulhon}, \citenamefont {Lasota}, \citenamefont {Paumard},\ and\ \citenamefont {Perrin}}]{Vincent:2020dij}%
  \BibitemOpen
  \bibfield  {author} {\bibinfo {author} {\bibfnamefont {F.~H.}\ \bibnamefont {Vincent}}, \bibinfo {author} {\bibfnamefont {M.}~\bibnamefont {Wielgus}}, \bibinfo {author} {\bibfnamefont {M.~A.}\ \bibnamefont {Abramowicz}}, \bibinfo {author} {\bibfnamefont {E.}~\bibnamefont {Gourgoulhon}}, \bibinfo {author} {\bibfnamefont {J.~P.}\ \bibnamefont {Lasota}}, \bibinfo {author} {\bibfnamefont {T.}~\bibnamefont {Paumard}}, \ and\ \bibinfo {author} {\bibfnamefont {G.}~\bibnamefont {Perrin}},\ }\href {\doibase 10.1051/0004-6361/202037787} {\bibfield  {journal} {\bibinfo  {journal} {Astron. Astrophys.}\ }\textbf {\bibinfo {volume} {646}},\ \bibinfo {pages} {A37} (\bibinfo {year} {2021})},\ \Eprint {http://arxiv.org/abs/2002.09226} {arXiv:2002.09226 [gr-qc]} \BibitemShut {NoStop}%
\bibitem [{\citenamefont {Akiyama}\ \emph {et~al.}(2022{\natexlab{c}})\citenamefont {Akiyama} \emph {et~al.}}]{EventHorizonTelescope:2022exc}%
  \BibitemOpen
  \bibfield  {author} {\bibinfo {author} {\bibfnamefont {K.}~\bibnamefont {Akiyama}} \emph {et~al.} (\bibinfo {collaboration} {Event Horizon Telescope}),\ }\href {\doibase 10.3847/2041-8213/ac6736} {\bibfield  {journal} {\bibinfo  {journal} {Astrophys. J. Lett.}\ }\textbf {\bibinfo {volume} {930}},\ \bibinfo {pages} {L15} (\bibinfo {year} {2022}{\natexlab{c}})}\BibitemShut {NoStop}%
\bibitem [{\citenamefont {Akiyama}\ \emph {et~al.}(2022{\natexlab{d}})\citenamefont {Akiyama} \emph {et~al.}}]{EventHorizonTelescope:2022urf}%
  \BibitemOpen
  \bibfield  {author} {\bibinfo {author} {\bibfnamefont {K.}~\bibnamefont {Akiyama}} \emph {et~al.} (\bibinfo {collaboration} {Event Horizon Telescope}),\ }\href {\doibase 10.3847/2041-8213/ac6672} {\bibfield  {journal} {\bibinfo  {journal} {Astrophys. J. Lett.}\ }\textbf {\bibinfo {volume} {930}},\ \bibinfo {pages} {L16} (\bibinfo {year} {2022}{\natexlab{d}})}\BibitemShut {NoStop}%
\bibitem [{\citenamefont {Akiyama}\ \emph {et~al.}(2022{\natexlab{e}})\citenamefont {Akiyama} \emph {et~al.}}]{EventHorizonTelescope:2022vjs}%
  \BibitemOpen
  \bibfield  {author} {\bibinfo {author} {\bibfnamefont {K.}~\bibnamefont {Akiyama}} \emph {et~al.} (\bibinfo {collaboration} {Event Horizon Telescope}),\ }\href {\doibase 10.3847/2041-8213/ac6675} {\bibfield  {journal} {\bibinfo  {journal} {Astrophys. J. Lett.}\ }\textbf {\bibinfo {volume} {930}},\ \bibinfo {pages} {L13} (\bibinfo {year} {2022}{\natexlab{e}})}\BibitemShut {NoStop}%
\bibitem [{\citenamefont {Akiyama}\ \emph {et~al.}(2022{\natexlab{f}})\citenamefont {Akiyama} \emph {et~al.}}]{EventHorizonTelescope:2022wok}%
  \BibitemOpen
  \bibfield  {author} {\bibinfo {author} {\bibfnamefont {K.}~\bibnamefont {Akiyama}} \emph {et~al.} (\bibinfo {collaboration} {Event Horizon Telescope}),\ }\href {\doibase 10.3847/2041-8213/ac6429} {\bibfield  {journal} {\bibinfo  {journal} {Astrophys. J. Lett.}\ }\textbf {\bibinfo {volume} {930}},\ \bibinfo {pages} {L14} (\bibinfo {year} {2022}{\natexlab{f}})}\BibitemShut {NoStop}%
\bibitem [{\citenamefont {Ghez}\ \emph {et~al.}(2003)\citenamefont {Ghez} \emph {et~al.}}]{Ghez:2003rt}%
  \BibitemOpen
  \bibfield  {author} {\bibinfo {author} {\bibfnamefont {A.~M.}\ \bibnamefont {Ghez}} \emph {et~al.},\ }\href {\doibase 10.1086/374804} {\bibfield  {journal} {\bibinfo  {journal} {Astrophys. J. Lett.}\ }\textbf {\bibinfo {volume} {586}},\ \bibinfo {pages} {L127} (\bibinfo {year} {2003})},\ \Eprint {http://arxiv.org/abs/astro-ph/0302299} {arXiv:astro-ph/0302299} \BibitemShut {NoStop}%
\bibitem [{\citenamefont {Ghez}\ \emph {et~al.}(2008)\citenamefont {Ghez} \emph {et~al.}}]{Ghez:2008ms}%
  \BibitemOpen
  \bibfield  {author} {\bibinfo {author} {\bibfnamefont {A.~M.}\ \bibnamefont {Ghez}} \emph {et~al.},\ }\href {\doibase 10.1086/592738} {\bibfield  {journal} {\bibinfo  {journal} {Astrophys. J.}\ }\textbf {\bibinfo {volume} {689}},\ \bibinfo {pages} {1044} (\bibinfo {year} {2008})},\ \Eprint {http://arxiv.org/abs/0808.2870} {arXiv:0808.2870 [astro-ph]} \BibitemShut {NoStop}%
%%CITATION = ARXIV:0808.2870;%%
\bibitem [{\citenamefont {Gillessen}\ \emph {et~al.}(2009)\citenamefont {Gillessen}, \citenamefont {Eisenhauer}, \citenamefont {Trippe}, \citenamefont {Alexander}, \citenamefont {Genzel}, \citenamefont {Martins},\ and\ \citenamefont {Ott}}]{Gillessen:2008qv}%
  \BibitemOpen
  \bibfield  {author} {\bibinfo {author} {\bibfnamefont {S.}~\bibnamefont {Gillessen}}, \bibinfo {author} {\bibfnamefont {F.}~\bibnamefont {Eisenhauer}}, \bibinfo {author} {\bibfnamefont {S.}~\bibnamefont {Trippe}}, \bibinfo {author} {\bibfnamefont {T.}~\bibnamefont {Alexander}}, \bibinfo {author} {\bibfnamefont {R.}~\bibnamefont {Genzel}}, \bibinfo {author} {\bibfnamefont {F.}~\bibnamefont {Martins}}, \ and\ \bibinfo {author} {\bibfnamefont {T.}~\bibnamefont {Ott}},\ }\href {\doibase 10.1088/0004-637X/692/2/1075} {\bibfield  {journal} {\bibinfo  {journal} {Astrophys. J.}\ }\textbf {\bibinfo {volume} {692}},\ \bibinfo {pages} {1075} (\bibinfo {year} {2009})},\ \Eprint {http://arxiv.org/abs/0810.4674} {arXiv:0810.4674 [astro-ph]} \BibitemShut {NoStop}%
%%CITATION = ARXIV:0810.4674;%%
\bibitem [{\citenamefont {Fragione}\ and\ \citenamefont {Loeb}(2020)}]{Fragione:2020khu}%
  \BibitemOpen
  \bibfield  {author} {\bibinfo {author} {\bibfnamefont {G.}~\bibnamefont {Fragione}}\ and\ \bibinfo {author} {\bibfnamefont {A.}~\bibnamefont {Loeb}},\ }\href {\doibase 10.3847/2041-8213/abb9b4} {\bibfield  {journal} {\bibinfo  {journal} {Astrophys. J. Lett.}\ }\textbf {\bibinfo {volume} {901}},\ \bibinfo {pages} {L32} (\bibinfo {year} {2020})},\ \Eprint {http://arxiv.org/abs/2008.11734} {arXiv:2008.11734 [astro-ph.GA]} \BibitemShut {NoStop}%
\end{thebibliography}%
\bibliographystyle{apsrev4-1}
\end{document}